\documentclass{article} 
\usepackage{iclr2022_conference,times}

\usepackage{microtype}
\usepackage{graphicx}
\usepackage{booktabs} 
\usepackage{amsmath,amssymb}
\usepackage{enumitem}
\usepackage{comment}
\usepackage{bm}
\usepackage{bbm}
\usepackage{color}
\usepackage{hyperref}
\usepackage{url}
\usepackage[T1]{fontenc}
\usepackage{hyperref}
\usepackage{url}
\usepackage{siunitx,booktabs,subcaption,tabularx}
\usepackage[margin=1in]{geometry}

\title{Fixed Points in Cyber Space: Rethinking \\ Optimal Evasion Attacks in the Age of AI-NIDS}
 
\author{Christian Schroeder de Witt\thanks{Corresponding author. Please contact at cs@robots.ox.ac.uk}  \\
Department of Engineering Science\\
University of Oxford\\
\texttt{cs@robots.ox.ac.uk} \\
\And
\hspace{-25.2em} Yongchao Huang\\
\hspace{-25.2em} Department of Computer Science \\
\hspace{-25.2em} University of Oxford \\
\hspace{-25.2em} \texttt{yongchao.huang@cs.ox.ac.uk} \\
\AND
Phil H.S. Torr \hspace{3em}\\
Department of Engineering Science\\
University of Oxford\\
\texttt{philip.torr@eng.ox.ac.uk} \\
\And
Martin Strohmeier \\
armasuisse Science and Technology \\
Thun, Switzerland\\
\texttt{martin.strohmeier@ar.admin.ch}
}

\newcommand{\Set}[1]{{\mathcal{#1}}}

\newcommand{\E}[0]{\mathbb{E}}
\renewcommand{\vec}[1]{{\bm{#1}}}

\newcommand{\methodone}[0]{FAST}
\newcommand{\methodtwo}[0]{CyberMARL}

\iclrfinalcopy 
\begin{document}

\maketitle

\begin{abstract}
Cyber attacks are increasing in volume, frequency, and complexity. In response, the security community is looking toward fully automating cyber defense systems using machine learning. However, so far the resultant effects on the coevolutionary dynamics of attackers and defenders have not been examined.
In this whitepaper, we hypothesise that increased automation on both sides will accelerate the coevolutionary cycle, thus begging the question of whether there are any resultant fixed points, and how they are characterised. Working within the threat model of Locked Shields, Europe's largest cyberdefense exercise, we study blackbox adversarial attacks on network classifiers. Given already existing attack capabilities, we question the utility of optimal evasion attack frameworks based on minimal evasion distances. Instead, we suggest a novel reinforcement learning setting that can be used to efficiently generate arbitrary adversarial perturbations. We then argue that attacker-defender fixed points are themselves general-sum games with complex phase transitions, and introduce a temporally extended multi-agent reinforcement learning framework in which the resultant dynamics can be studied. We hypothesise that one plausible fixed point of AI-NIDS may be a scenario where the defense strategy relies heavily on whitelisted feature flow subspaces. Finally, we demonstrate that a continual learning approach is required to study attacker-defender dynamics in temporally extended general-sum games.

\end{abstract}

\section{Introduction}

Driven by ever-increasing infrastructure growth, cyber attacks are becoming increasingly frequent, sophisticated, and concealed. The use of semi-automated network intrusion detection systems using machine learning technology (ML-NIDS) reduces the workload on human network operators. However, modern cyber attacks are in turn increasingly drawing on machine learning techniques, thus perpetuating the continual arms race between cyber attackers and defenders. 

To mitigate the natural asymmetry between attackers and defenders, recent work \citep{meier_towards_2021} has, possibly for the first time, sought to described a fully-automated network defense system. We will in the following refer to such systems as \textit{AI-NIDS}, in order to distinguish human-in-the-loop systems relying on machine learning technology (\textit{ML-NIDS}) from systems that make decisions in a fully autonomous fashion.

While the concept of AI-NIDS may seem intriguing to practitioners, to the best of our knowledge, so far nobody has studied their effect on the coevolutionary attacker-defender dynamics. This is surprising given the potentially game-changing implications of interacting, fully automated cyber attack and defense systems in the near future: While the speed of today's cyber security arms race is mostly limited by human ingenuity, automated systems of the future will interact and coevolve on significantly smaller timescales, learning new exploits and defenses against over the course of not months, but hours, or even milliseconds. Analogous escalations due to increased automation are starting to be observed in financial markets \citep{alexandrov_exploring_2021}.

In this paper, we argue that the advent of AI-NIDS may trigger a coevolutionary explosion with wide-reaching consequences for cyber security practice. 
In this new era of nearly-instantaneous attacker-defender coevolution, we anticipate that the study of mutual best responses between fixed populations of attack and defense techniques will have to be replaced by a better understanding of the emerging dynamical fixed points that the attacker-defender systems might evolve toward.

Without loss of generality, we in this work focus on system dynamics arising from network-classifier based defenses against botnet infiltration. Botnets pose an increasingly severe security threat to the global security environment, enabling crimes such as information theft and distributed denial-of-service attacks \citet{abu_rajab_multifaceted_2006}. Network flow classification (NFC) is a central security component in botnet detection systems \citep{abraham_comparison_2018}. In traditional network intrusion detection systems (IDS), classifier rules are little more than codified expert knowledge based on a set of known malicious and benign behaviour. With the rapid growth of networking systems, ever-evolving attack patterns and the rising complexity of communication protocols, state-of-the-art NFC techniques increasingly rely on machine learning (ML) to train parameterised classifiers directly from large amounts of labeled examples \citep{lakhina_diagnosing_2004}. In fact, the automated cyber defense system proposed by \citet{meier_towards_2021} makes heavy use of machine learning-based NFCs.\\

While being promising in many practical applications \citep{wagner_mimicry_2002}, it has been shown that ML-based classifiers are inherently vulnerable to adversarial attacks \citep{dalvi_adversarial_2004}. One such class of attacks includes the poisoning of training data pools \citep{barreno_can_2006}. Another such class of attacks are optimal evasion attacks \citep{nelson_near-optimal_2010}, in which an attacker changes its communication patterns such as to avoid NFC detection. In this paper, we restrict ourselves to the study of so-called \textit{blackbox} optimal evasion attacks, referring to settings in which attackers and defenders do not have access to each other's internal models but can solely infer the other's state through real-world interactions. Adversarial blackbox attacks on network flow classifiers have recently been demonstrated using reinforcement learning with a sparse feedback signal in order to generate adversarial perturbations allowing malicious communication to be masked \citep{apruzzese_deep_2020}.\\

 While being ubiquitously studied due their satisfying mathematical properties, optimal evasion frameworks generally neglect a crucial characteristic of real-world NFC evasion attacks. In fact, radically altering malicious traffic distributions in stochastic flow feature space does not necessarily impose any tangible costs on the attacker \footnote{This observation was made already in \cite{merkli_evaluating_2020}, but remained largely inconsequential.}: An abundance of specialised networking software \cite[Scapy]{community_scapy_2021}, compression, encryption, and myriads of openly available repositories of wildly diverse C\&C communication protocols\footnote{Try \url{https://www.google.com/search?q=github+command+and+control+protocols}.} mean that attackers do not need to search for the smallest possible adversarial perturbations. Instead, as long as overall information throughput is guaranteed to remain above a certain operational threshold - that is usually small compared to benign network traffic -, we hypothesise that real-world  attackers can traverse statistical flow feature (SFF) space almost arbitrarily at little cost. With the advent of increasingly powerful AI-based program synthesis methods \citep{church_application_1963, waldinger_prow_1969, johnson_inferring_2017, parisotto_neuro-symbolic_2016, devlin_robustfill_2017}, we expect that future attackers will be able to generate efficient C\&C communication protocols with associated statistical flow feature properties \textit{a la carte}, and at negligible cost.

The framework introduced by \citet{apruzzese_deep_2020} does not readily extend to settings in which adversarial perturbations of large or arbitrary sizes need to be generated as each feature modification requires a full environment step. Very long episodes are not only inefficient to generate, but can also cause issues with temporal credit assignment during training. We thus introduce \textit{Fast Adversarial Sample Training} (\methodone), a reinforcement learning method for blackbox adversarial attacks that can generate continuous perturbations of arbitrary size in just a single step. We show that \methodone~can be efficiently used to generate discrete features. For completeness, we show \methodone~to be efficient and easily tunable even in minimal evasion distance settings. 

Defense against NFC blackbox attacks is traditionally achieved by \textit{hardening}, i.e. by supervised training on known (or artificially generated) malicious perturbations. In fact, within the \citet{apruzzese_deep_2020} framework, we demonstrate that, under favourable conditions, NFCs can be hardened sufficiently in just one retraining step. This means that subsequent attacks using the same method will largely fail to deceive the hardened classifier within over the same flow sample distribution. We thus present empirical evidence that, under certain conditions, blackbox minimal distance evasion attack dynamics, can indeed converge to a no-attack fixed point. However, once the minimal evasion distance framework is abandoned, such fixed points are clearly unlikely to be stable.

Instead, we argue that understanding the dynamics of \textit{arbitrary distance evasion attacks} (ADEA) requires modeling the attacker-defender system as a \textit{temporally extended general-sum game}. We note that the game is not \textit{zero-sum}, as may commonly be assumed, because, while attackers try to infect nodes and defenders try to keep from being infected, both attackers and defenders are simultaneously interest in keeping network services operational: If an attacker compromises network operations to a noticeable extent, then the risk of security escalations or even network shutdowns in response increases - this in turn compromises botnet operations. In return, defenders might worry less about possible infections as long as network operations remain unaffected. The goals of attackers and defenders are thus partially aligned. We proceed to formulate such games as a deep multi-agent reinforcement learning problem, \methodtwo, in which both attackers and defenders share the same network channel. \methodtwo~ easily accommodates for complications found in real-world NIDS operations, such as delayed feedbacks \citep{apruzzese_modeling_2021}. In addition, in \methodtwo, we explicitly assume NFCs to operate concurrently to other ML-NIDS techniques, such as anomaly detectors and blacklisting. We illustrate the utility of \methodtwo~ in modeling a \textit{mode transition}, i.e. a period of time during which the attacker suddenly changes attack distributions, thus pushing the defender NFC temporarily out-of-distribution. 

How could cyber defense systems of the distant future defend themselves effectively from ADEAs? We hypothesise that one plausible fixed point of attacker-defender coevolution may be a scenario we dub \textit{Whitelisting Hell}. In \textit{Whitelisting Hell}, AI-NIDS automatically drop network traffic whose SFFs lie outside of very narrow, temporally changing, whitelisted subspaces (or "corridors"). In order to be able to create and operate a botnet under such conditions, the attacker needs to predict the dynamics of the whitelisted subspaces in order to generate C\&C flows within these. 

Whether in \textit{Whitelisting Hell}, or under the more general ADEA conditions explored by \methodtwo, possible attack strategies may seek to exploit a weakness of supervised classifiers: Neural networks, as well as other classifiers, suffer from catastrophic forgetting. As an antidote, training samples could be exhaustively stored and the classifier be retrained continuously, however, this may not be possible given traffic volumes, as well as real-time constraints and computation budgets even of large-scale AI-NIDS systems. We suggest and empirically illustrate that NFC defenses in future AI-NIDS may benefit from \textit{continual learning} \citep{parisi_continual_2019} techniques.

This paper proceeds by first providing the necessary background on botnet detection, network classifiers, Locked Shields, blackbox adversarial attacks, and deep multi-agent reinforcement learning (Section \ref{sec:back}). We subsequently introduce and evaluate \methodone~(Section \ref{sec:cont}), followed by \methodtwo~(Section \ref{sec:cybermarl}), and conclude with a description of \textit{Whitelisting Hell}, and an empirical illustration of the demonstration of continual learning techniques in temporally extended attacker-defender settings (Section \ref{sec:white}).



\section{Background}
\label{sec:back}

\subsection{Botnets, Network Flow Classifiers and Locked Shields}

\begin{figure}[h]
\centering
        \includegraphics[width=0.7\linewidth]{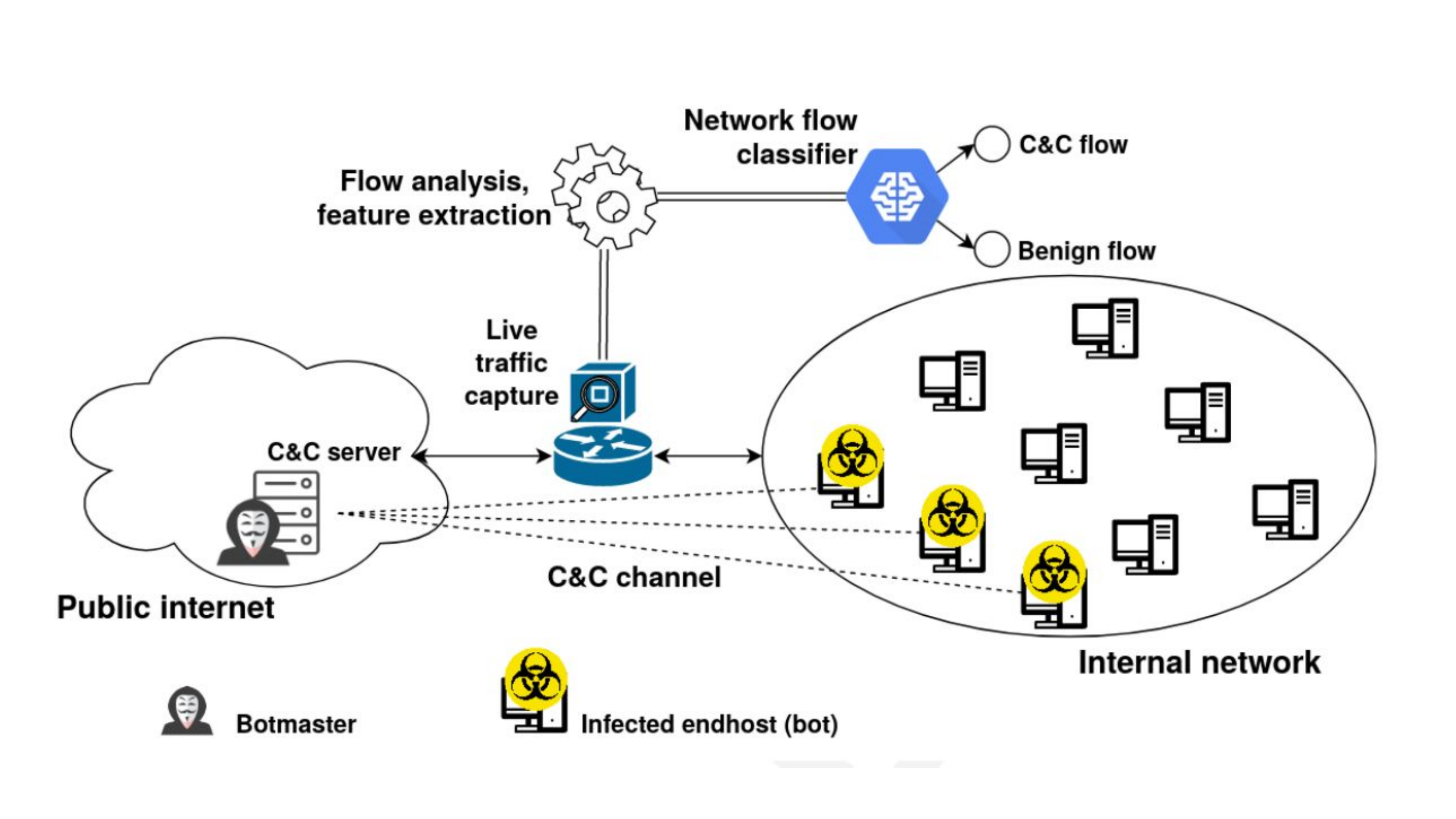}
        \caption{Network flow classification for detection of C\&C channels. Taken from \citet{merkli_evaluating_2020}}
\label{fig:netflow}
\end{figure}

Botnets pose an increasingly severe security threat to the global security environment, enabling crimes such as information theft and distributed denial-of-service attacks \citep{abu_rajab_multifaceted_2006}.
Network flow classification (NFC) is a central security component in botnet detection systems \citep{abraham_comparison_2018}. In NFC, a blue team classifier decides whether a given intercepted network package is benign or part of team red's botnet communication stream, for example the communication between an infected host and a command and control (C\&C) server (see Figure \ref{fig:netflow}). Packages that are identified as malicious may simply be dropped by team blue, thus blocking team red's C\&C channels. \\

In traditional network intrusion detection systems (IDS), classifier rules codified expert knowledge based on a set of known malicious and benign behaviour. With the rapid growth of networking systems, ever-evolving attack patterns and the rising complexity of communication protocols, state-of-the-art NFC techniques increasingly rely on machine learning (ML) to train parameterised classifiers directly from large amounts of labeled examples \citep{lakhina_diagnosing_2004}. \\

While being promising in many practical applications \citep{wagner_mimicry_2002}, it has been shown that ML-based classifiers are inherently vulnerable to adversarial attacks \citep{dalvi_adversarial_2004}. One such class of attacks includes the poisoning of training data pools \citep{barreno_can_2006}. Another such class of attacks are optimal evasion attacks \citep{nelson_near-optimal_2010}, in which an attacker changes its communication patterns such as to avoid NFC detection.\\

Locked Shields (LS) is one of the world's largest cyber defense exercises, being held annually in Tallinn, Estonia \citep{ccdcoe_locked_2020}. In LS, Team Red (R) attempts to infiltrate internal network nodes belonging to Team Blue (B) in order to establish a botnet. While team red is supplied with many details about the network and allowed to infiltrate it ahead of time, team blue only has band-width limited access to a few nodes and tight computational budget constraints. Team Blue's goal consists of identifying infiltrated nodes and interrupt the operation of Team Red's botnet.

\paragraph{Statistical Flow Features}
Statistical flow features (SFF), such as those extracted by CICFlowMeter \citep{cicflowmeter_2020}, are based entirely on package meta-information, not content, meaning they can be evaluated even for encrypted or compressed flows. Despite - or indeed because of - this level of abstraction, SFFs have been found to be useful features for a variety of cyber security tasks.

\paragraph{Adversarial attacks and defenses}
Deep Neural Networks (DNNs) are known to be vulnerable to adversarial perturbations \citep{ren_adversarial_2020}. An \textit{adversarial perturbation} $\delta$ is a small term that, when added to a benign sample $x$ with classifier output label $y$, results in an adversarial sample that is both realistic, but results either in a different classifier output label (untargeted) or a maliciously chosen one (targeted). Many methods for generating such adversarial samples try to induce visual realism by minimizing a pixel-space $L_p$ norm between the adversarial sample and the benign sample \citep{carlini_towards_2017,liu_delving_2017}. However, such a realism criterion is problematic in flow feature space \citep{merkli_evaluating_2020}.

\paragraph{Optimal evasion attacks}

\begin{figure}[h]
\centering
    \begin{minipage}{0.43\textwidth}
        \includegraphics[width=0.8\textwidth]{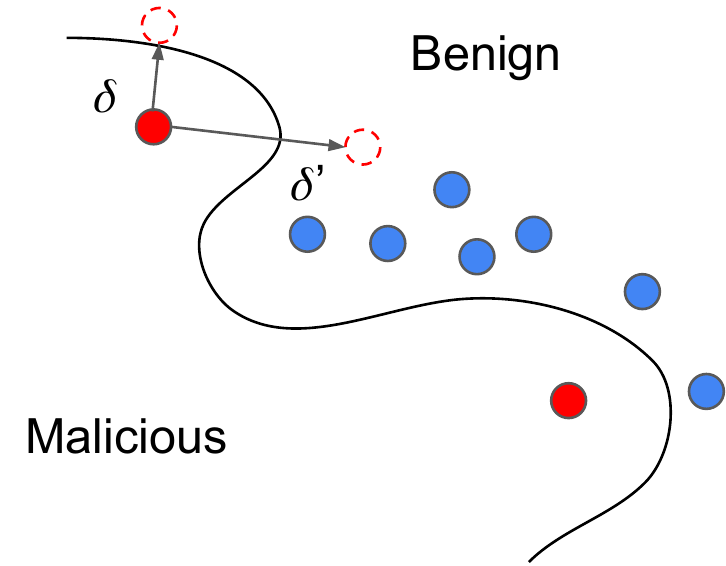}
        \caption{An illustration of minimal distance evasion attacks.}
        \label{fig:minev}
    \end{minipage}
    \begin{minipage}{0.55\textwidth}
        \centering
         \includegraphics[width=0.8\textwidth]{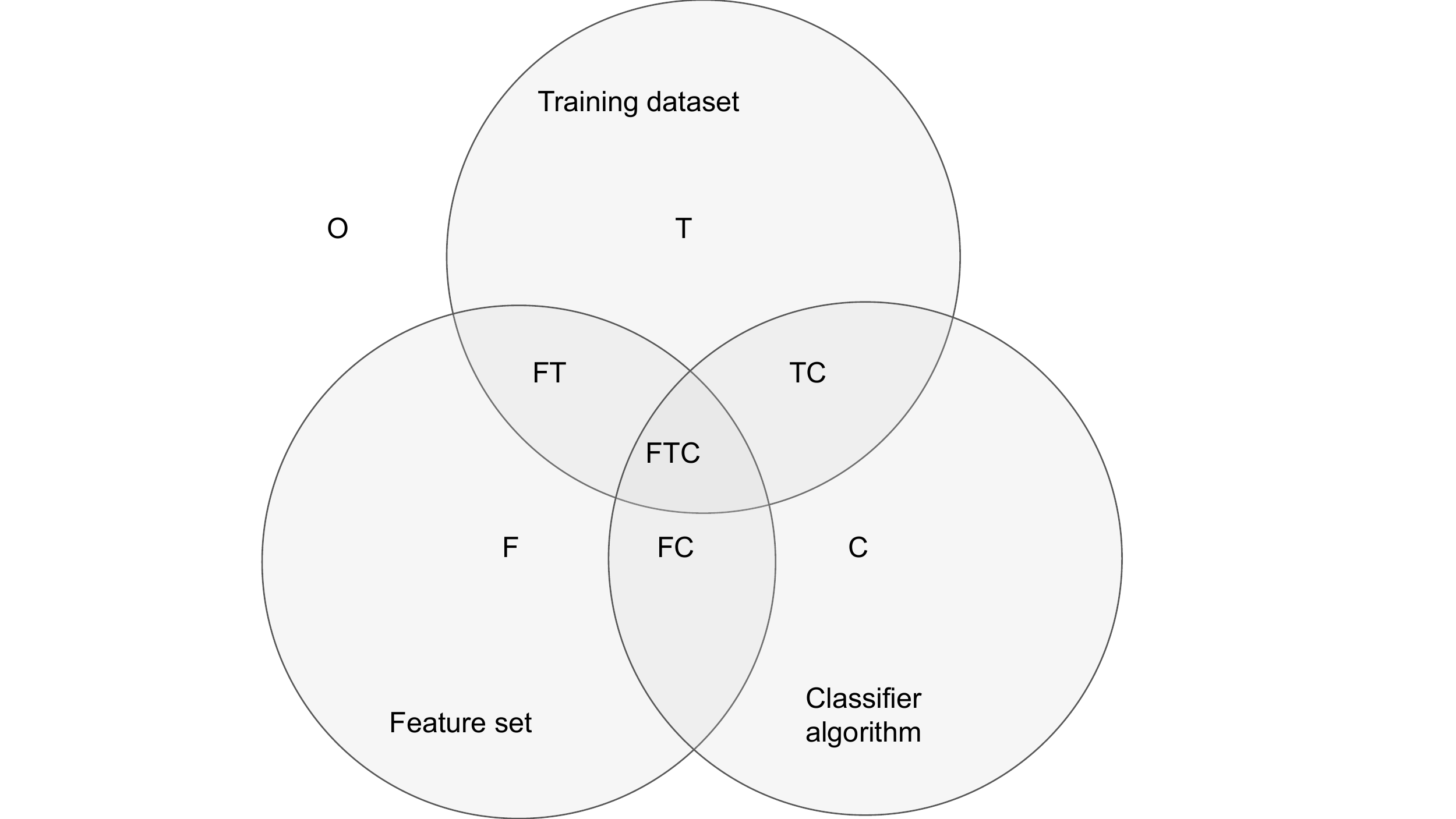}
        \caption{Levels of access. Taken from \citep{merkli_evaluating_2020}.}
        \label{fig:access}
    \end{minipage}
\end{figure}

An \textit{optimal evasion} attack is an adversarial attack using a perturbation that is optimal with respect to some optimality criteria. Although in principle, this optimality criterion could be chosen freely, recent work has almost exclusively focused on evasion attacks where the size of the adversarial perturbation generated is to be minimised: in other words, attackers are assumed to incur a cost proportional to the size of the adversarial perturbation. Formally,  such \emph{minimal distance evasion attacks (MDEA), see Figure \ref{fig:minev}} \citep{merkli_evaluating_2020} posit that, given some binary discriminator $D_{\text{rf}}$, the goal is to identify perturbations $\delta x_i$ such that $D_{\text{rf}}(x_i+\delta x_i)=\neg D_{\text{rf}}(x_i)$ while minimizing $\left\|\delta x_i\right\|_p^2,\ p\geq 1$. \citet{merkli_evaluating_2020} employ a random forest discriminator and find $\delta x_i$ by solving a MILP problem first proposed by \citep{kantchelian_evasion_2016}, using an apriori fixed set of $N$ labeled samples $x_i\in\mathcal{X}$.

\paragraph{Blackbox attacks}
Adversarial attacks may occur with various levels of access (see Figure \ref{fig:access}). Attackers might, for example, have access to the defender's NFC model weights, and/or the classifier features, or the defender's training set. In this paper, we focus on so-called \textit{blackbox} settings in which the attacker cannot be assumed to have access to any of these 


\subsection{Reinforcement Learning and General-Sum Games}

\paragraph{Partially-Observable Stochastic Games}

\citep[POSGs][]{hansen_dynamic_2004} describe multi-agent tasks where a number of agents with individual goals choose sequential actions under partial observability and environment stochasticity.
POSGs can be formally defined by a tuple $\langle \mathcal{N} , \Set S, \Set U, P, r_a, \Set Z, O, \rho, \gamma \rangle$. 
Here $s \in \Set S$ describes the state of the environment, discrete or continuous, and $ \mathcal{N} := \{1,\dots,N\}$ denotes the set of $N$ agents. $s_0 \sim \rho$, the initial state, is drawn from distribution $\rho$. At each time step $t$, all agents $a \in \mathcal{N}$ simultaneously 
choose actions $u^a_t \in \Set U$ which may be discrete or continuous. This yields the joint action $\vec{u}_t := \{u_t^a\}_{a=1}^{N} \in \Set U^N$. 
The next state $s_{t+1} \sim P(s_t, \vec u_t)$ is drawn from transition kernel $P$ after executing the joint action $\vec u_t$ in state $s_t$. Subsequently, agent $a$ receive a scalar reward $r^a_t = r^a(s_t,  u^a_t)$.

Instead of being able to observe the full state $s_t$, in a POSG each agent $a \in \mathcal{N}$ can only draw an individual 
local observation $z_t^a \in \Set Z, \vec z_t := \{z_t^a\}_{a=1}^N$,  
from the observation kernel $O(s_t, a)$. The history of an agent's observations and actions is denoted by
$\tau^a_t \in \Set T_t := (\Set Z \times \Set U)^t \times \Set Z$.
The set of all agents' histories is given by $\vec\tau_t := \{\tau^a_t\}_{a=1}^N$.
Each agent $a$ chooses its actions with a decentralised 
policy $u^a_t \sim \pi^a(\cdot|\tau^a_t)$ that is
based only on its individual history. 

Each agent attempts to learn 
an individual {\em policy} $\pi^a(u^a|\tau^a_t)$
that maximises the agent's expected discounted return, 
$J(\pi^a) \doteq \E[\sum_{t=0}^{\infty} \gamma^t r^a_t]$, where $\gamma \in [0, 1)$ is a discount factor. 
$\pi^a(u^a|\tau^a_t)$ induces an individual action-value function $Q^\pi_a:=\mathbb{E}\left[\sum_{i=0}^{\infty}\gamma^ir^a_{t+i}\right]$
that estimates the expected discounted return of agent $a$'s action $u^a_t$ 
in state $s_t$ with agent history $\vec \tau_t$.

\paragraph{Centralised learning with decentralised execution.}
\label{sec:ctde}
Reinforcement learning policies can often be learnt in simulation or in a laboratory. In this case, on top of their local observation histories, agents may have access to the full environment state and share each other's policies and experiences during training. The framework of \textit{centralised training with decentralised execution (CTDE)} \citep{oliehoek_concise_2016,kraemer_multi-agent_2016} formalises the use of centralised information to facilitate the training of decentralisable policies.




\paragraph{Deep Q-learning}

{\em Deep $Q$-Network} (DQN) \citep{mnih_human-level_2015} uses a deep neural network to estimate the action-value function, $Q(s, \vec\tau, \vec u; \theta) 
\approx \max_\pi Q^\pi(s, \vec\tau, \vec u)$, 
where $\theta$ are the parameters of the network. 
For the sake of simplicity, 
we assume here feed-forward networks,
which condition on the last observations $\vec z^a_t$,
rather than the entire agent histories $\vec \tau_t$.
The network parameters $\theta$ are trained by gradient descent on the mean squared regression loss:
\begin{align} \label{eq:dqn_loss}
	\mathcal{L}^{\scriptscriptstyle\text{DQN}}_{[\theta]} & :=  
	\E_{\Set D}\Big[ \big( y_t^{\scriptscriptstyle\text{DQN}} 
	- Q(s_t, \vec z_t, \vec u_t; \theta) \big)^2 \Big],     
\end{align} 
where $y_t^{\scriptscriptstyle\text{DQN}} := r_t + \gamma \max_{\vec u'}Q(s_{t+1}, \vec z_{t+1}, \vec u';\theta^{-})$ and
 the expectation is estimated with transitions  
$(s_t, \vec z_t, \vec u_t, r_t, s_{t+1}, \vec z_{t+1}) \sim \Set D$ 
sampled from an {\em experience replay buffer} $\Set D$ \citep{lin_self-improving_1992}. 
The use of replay buffer reduces correlations 
in the observation sequence. 
To further stabilise learning,
$\theta^{-}$ denotes parameters of a {\em target network} 
that are only periodically copied from the most recent $\theta$. While we do not employ DQN in this paper, it is an important reference algorithm in reinforcement learning \citep{sutton_reinforcement_2018}. Independent Q-learning (IQL) 
is a simple extension of single-agent Q-learning to multi-agent settings \citep{tan_multi-agent_1993}. 

\paragraph{Deep Deterministic Policy Gradients (DDPG)}
\label{sec:ddpg}
Deep Deterministic Policy Gradient (DDPG) \citep{lillicrap_continuous_2015} is an actor-critic algorithm that poses an important alternative to continuous Q-learning \citep{gu_continuous_2016}. In DDPG, an actor has a deterministic policy $\mu$ parametrised by $\theta$. The actor is learnt alongside a critic, $Q$ that conditions on the agent's observation (or the full environment state $s$, if centralised training is available). The critic is updated using a TD-error loss:
\begin{align}	 
	\mathcal{L}^{\scriptscriptstyle\text{DPG}}_{[\phi]} &:= 
		\E_{\Set D}\Big[
		\Big(y_t
		- Q^{\mu}\!\big(s_t, \vec u_t; \phi \big)
		\Big)^{\!2} 
	\,\Big], 
\end{align} 
where $y_t := r_t + \gamma \, Q^{\mu}\big(s_{t+1}, \vec\mu(\vec\tau_{t+1}; \vec\theta') ; \phi'\big)$ and
transitions are sampled from a replay buffer $\Set D$ \citep{lin_self-improving_1992}
and $\vec\theta'$ and $\phi'$ are target-network parameters.
 
\paragraph{Multi-Agent Deep Deterministic Policy Gradient (MADDPG)}
\label{sec:maddpg}

{\em Multi-agent deep deterministic policy gradient} \citep[MADDPG][]{lowe_multi-agent_2017} is an actor-critic method that works in both cooperative and competitive MARL tasks with discrete or continuous action spaces. 
MADDPG was originally designed for the general case of {\em partially observable stochastic games} \citep{kuhn_extensive_1953}, in which it learns a separate actor and centralised critic for each agent such that agents can learn arbitrary reward functions - including conflicting rewards in competitive settings. In this paper, we employ a variant of MADDPG with continuous action spaces.
We assume each agent $a$ has a deterministic policy $\mu^a$, 
parameterised by $\theta^a$, with
$\vec\mu(\vec\tau;\vec\theta)\!:=\!\{\mu^a(\tau^a; \theta^a)\}_{a=1}^N$.
For POSGs, MADDPG learns individual centralised critics 
$Q_a^{\vec\mu}\left(s, \vec u; \phi\right)$ 
for each agent $a$ with shared weights $\phi$
that condition on the full state $s$ and the joint actions $\vec u$ of all agents. 
The policy gradient for $\theta^a$ is given by:
\begin{equation} \nonumber
\setlength{\abovedisplayskip}{4pt}\setlength{\belowdisplayskip}{4pt}
	\!\nabla_{\!\theta^a} \! \Set L^\mu_{[\theta^a]} \!:=
	-\E_{\Set D}\!\Big[\!\nabla_{\!\theta^a} \! \mu^a\!(\tau^a_t; \!\theta^a) 
			\nabla_{\!u^{\!a}} \! Q_a^{\vec\mu}\!(s_t, \!\hat{\vec u}^a_t; \!\phi)
			\!\big|_{u^{\!a}=\mu^{\!a}(\tau_t^a)}\! \Big]\!,
\end{equation}
where $\hat{\vec u_t^a} := \{u_t^1, \ldots, u_t^{a-1}, u^a, u_t^{a+1}, \ldots, u^N_t\}$ and 
 $s_t,\vec u_t, \vec \tau_t$ are sampled from a replay buffer $\Set D$.
The shared centralised critic $Q_{a}^{\vec\mu}$ is trained by minimising the following loss:
\begin{align}	 
	\mathcal{L}^{\scriptscriptstyle\text{DPG}}_{[\phi]} &:= 
		\E_{\Set D}\Big[
		\Big(y_t^a
		- Q_a^{\vec\mu}\!\big(s_t, \vec u_t; \phi \big)
		\Big)^{\!2} 
	\,\Big], 
\end{align} 
where $y_t^a := r_t + \gamma \, Q_a^{\vec\mu}\big(s_{t+1}, \vec\mu(\vec\tau_{t+1}; \vec\theta') ; \phi'\big)$ and
transitions are sampled from a replay buffer $\Set D$ \citep{lin_self-improving_1992}
and $\vec\theta'$ and $\phi'$ are target-network parameters.

\subsection{Blackbox Adversarial Attacks}


Network Flow Classifiers (NFCs) are mappings $C:\mathcal{F}\rightarrow[0,1]$ that take a sample $x\in\mathcal{F}$, where $\mathcal{F}$ is the feature space, to a score $\rho(x)$. Scores $\geq0.5$ indicate a malicious sample,  scores below indicate a benign sample. Given a sample $x_{adv}$ for which $C(x_{adv})<0.5$, the task of an adversarial sample generator is then to generate a perturbation $\delta$ such that $C(x_{adv}+\delta)\geq 0.5$ \footnote{Such a perturbation is likely not unique.}. Throughout this paper, $C$ corresponds to a fixed classifier trained according to section \ref{sec:cont} using a random class-balanced set of $10^6$ samples from $2018$ Locked Shields data.

While neural network architectures are starting to catch up \citep{chernikova_fence_2020, apruzzese_deep_2020}, state-of-the-art classifier models are generally based on random forests \citep{merkli_evaluating_2020}.
\citet{kantchelian_evasion_2016} introduce a specialised evasion attack generation scheme for random forests that requires the costly solution of a mixed integer linear program (MILP) for each adversarial sample. Not only greatly does this limit the rate of adversarial samples that can generated, but, in addition, the inclusion of side constraints - e.g. to ensure that the generated samples are realistic - is greatly limited. 
As random forests are not end-to-end differentiable, approaches employing generative adversarial schemes are not immediately applicable \citep{NIPS2014_5ca3e9b1}.

\begin{figure}[h]
\centering
\includegraphics[width=0.7\linewidth]{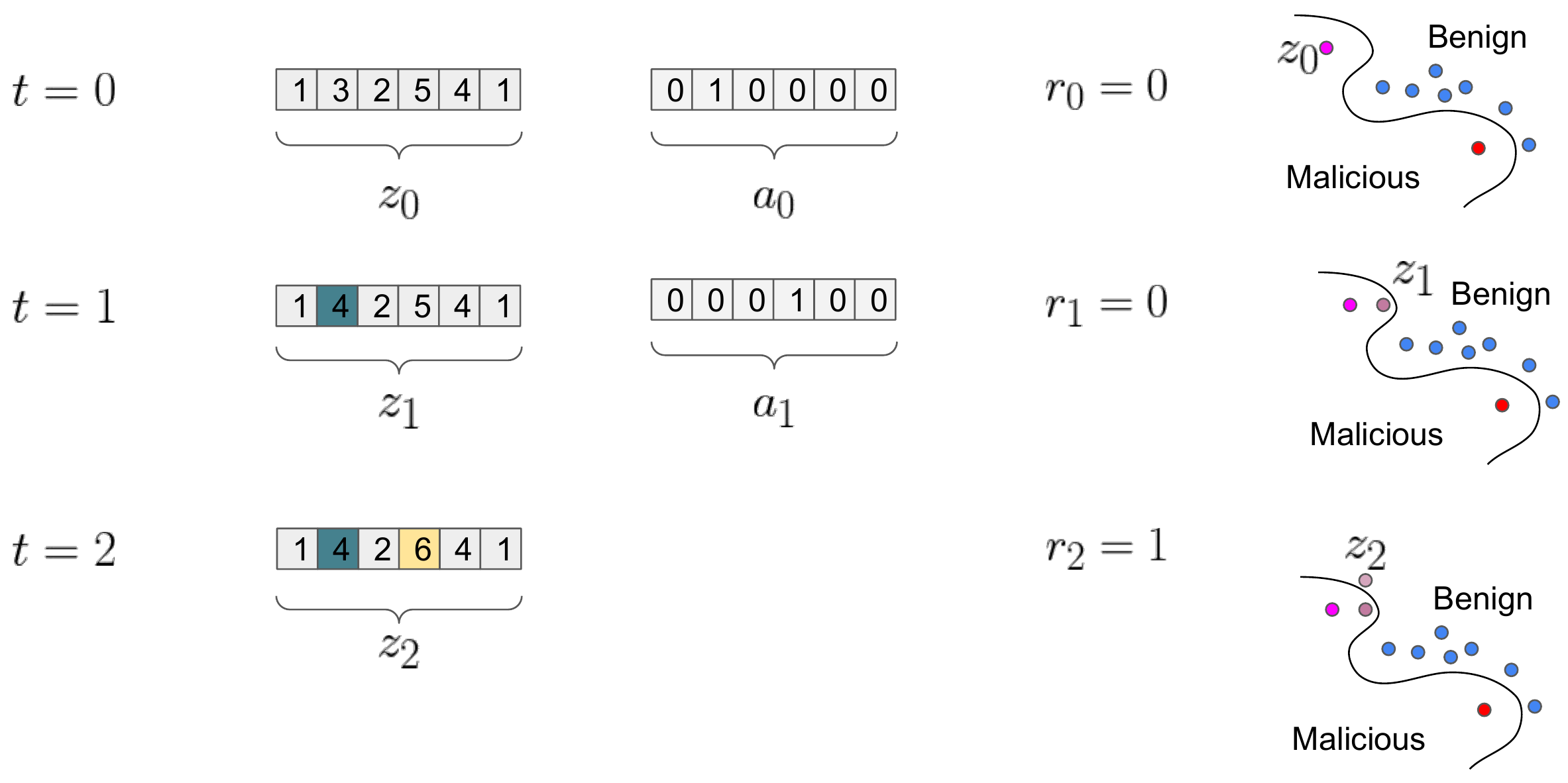}
\caption{An illustration of \citep{apruzzese_deep_2020}.}
\label{fig:apr}
\end{figure}

\citet{apruzzese_deep_2020} introduce a novel reinforcement learning-based method for generating adversarial perturbations in blackbox settings. Their method employs Q-learning over episodes with step-wise feature modifications (see Figure \ref{fig:apr}). While empirically shown to be able to generate efficient attacks, \citet{apruzzese_deep_2020}'s method suffers from potentially extremely long episode lengths if perturbation sizes are substantial, making it practically inapplicable outside of MDEA settings. Even within MDEA settings, the episodic nature of perturbation generation may impede the learning process as it may introduce temporal credit-assignment issues. 

\section{Optimal Evasion with Arbitrary Single-Step Perturbations}
\label{sec:cont}

To overcome the limitations of avdersarial sample generation in blackbox settings posed by \citet{apruzzese_deep_2020}'s method, we now introduce a novel setting in which, discrete or continuous, perturbations of arbitrary size are generated in a single step. 

\begin{figure}[h]
\begin{center}
\includegraphics[width=0.4\linewidth]{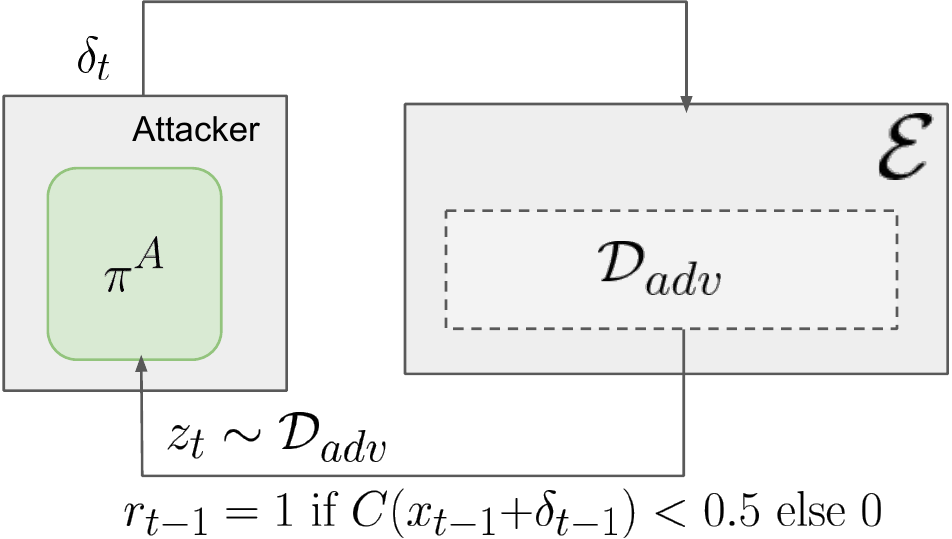}
\end{center}
\caption{Schematic depiction of \methodone's reinforcement learning loop, where $\mathcal{E}$ denotes the environment.} 
\label{fig:FAST}
\end{figure}

\paragraph{Fast Adversarial Sample Training (FAST).} (\methodone) formulates the process of finding a suitable $\delta$ as a single-agent reinforcement learning problem as follows (see Figure \ref{fig:FAST}): An attacker with policy $\pi^A$ receives an i.i.d. random sample $z\sim\mathcal{D}_{adv}$ from the environment $\mathcal{E}$. The attacker then constructs a perturbation $\delta:=\pi^A(z)$ and receives a reward that is $+1$ if the NFC has been fooled, i.e. $C(x+\delta) < 0.5$ and $0.0$ otherwise. In contrast to \citet{apruzzese_deep_2020}, we employ deep deterministic policy gradients (DDPG) (see Section \ref{sec:back}) and generate continuous perturbations, which are discretised post-hoc if feature spaces are discrete. Each episode terminates after a single step.

To illustrate \methodone's versatility even in MDEA settings, we evaluate \methodone~on a set of $10^6$ class-balanced i.i.d. sampled sets of Locked Shields samples $\mathcal{D}_{ben},\ \mathcal{D}_{mal}$ from the year 2018. We train the attacker policy $\pi$ using DDPG (see section \ref{sec:ddpg}), using three-layered fully connected neural network architecture with $\tanh$-activation functions and $64$ activations per layer for both actor and critic (see Figure \ref{fig:drl_policy}). We fix our actor learning rate at $10^{-4}$, and our critic learning rate at $10^{-3}$ and choose the discount factor $\gamma=0.99$. To incentivise small perturbations , we add a regulariser the L$2$ norm $\|\delta\|_2$. Note that perturbation constraints are feature-wise and scaled to feature-wise normalised features - absolute perturbations can be derived by first multiplying $\delta$ by the standard deviation of the respective feature and then adding its mean (see Table \ref{table:feat}). 

We empirically investigate the \textit{attack rate}, i.e. the ratio of generated perturbations that fool the classifier, after $10^6$ training steps. We find that \methodone~is able to fool the classifier about $15\%$ of the time even if we restrict feature-wise perturbations to $1/100$th of the feature standard deviation. If we increase the allowed perturbations by another factor $10$, then we reach almost perfect attack rates (see Figure \ref{fig:prob1_6_2}). We leave detailed empirical comparisons of \methodone~ with \citet{apruzzese_deep_2020}'s method for future work as our methodological advances are clear, and further empirical investigation lies beyond the scope of our paper.

\begin{table*}[h]
    \centering
    \begin{minipage}{1.0\textwidth}
        \centering
    \begin{tabular}{@{\extracolsep{5pt}} lccccc}
      \toprule
    Year \textbackslash $\|\delta\|$ & 0.001 & 0.003 & 0.01 & 0.03 & 0.1 \\ 
      \midrule
    2018 & 0.84\% & 1.2\% & 14.9\% & 63.6\% & 97.1\%\\
      \bottomrule
    \end{tabular}
        \caption{Attack rate for \methodone~given different hard upper constraints on the perturbation L$2$-norm $\|\delta\|_2$. The LightGBM model has been trained on the top $20$ most important features from class-balanced random samples from 2018 Locked Shields data. Note that the perturbation norm is expressed relative to feature-wise normalised samples.}
        \label{fig:prob1_6_2}
    \end{minipage}%
\end{table*}

\section{From Optimal Evasion Attacks To Temporally-Extended General-Sum Games}
\label{sec:cybermarl}


\subsection{Iterative Hardening}

\begin{figure*}[h]
\begin{center}
\includegraphics[width=0.8\linewidth]{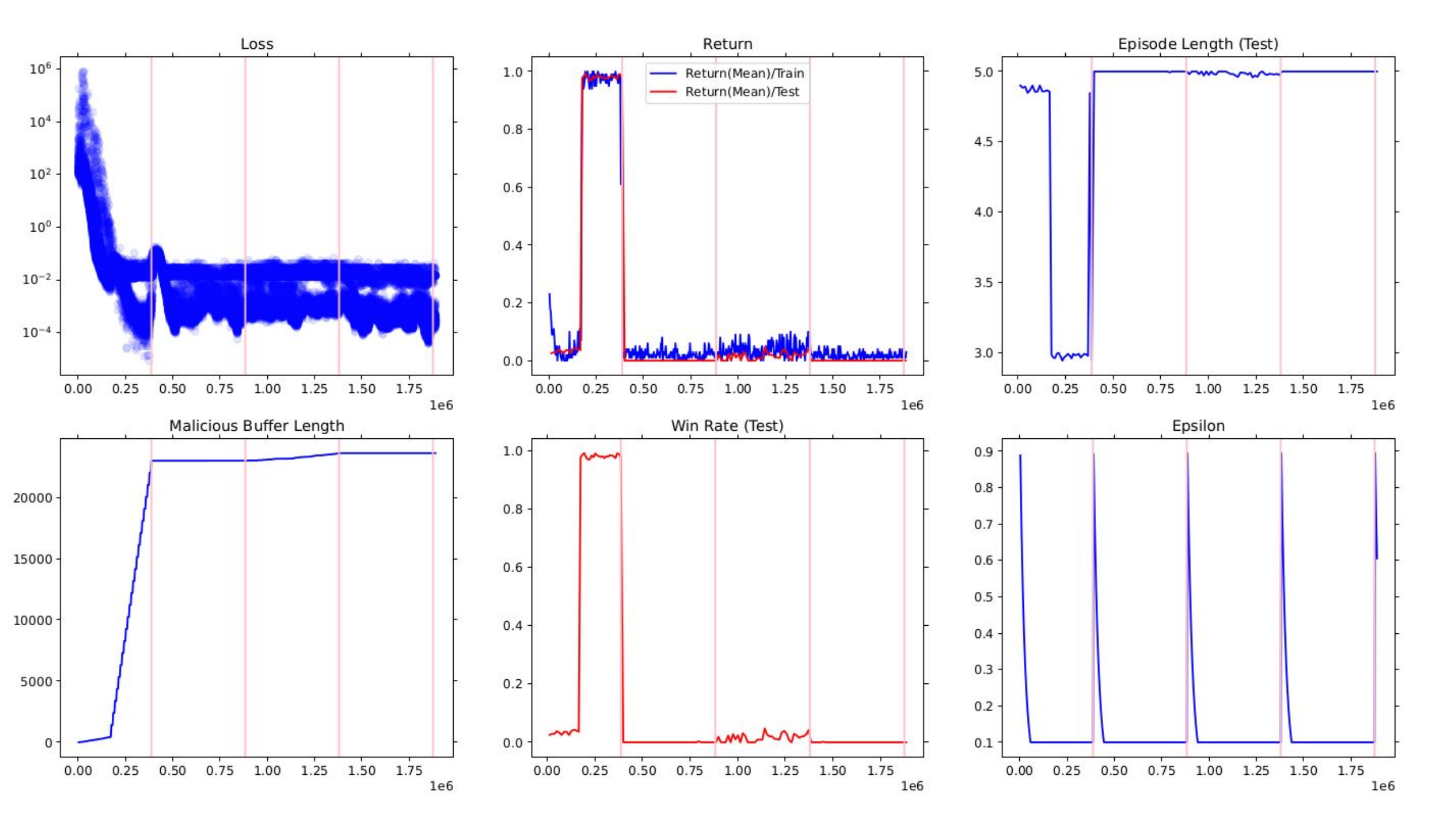}
\end{center}
\caption{Iterative Hardening in the Q-learning framework introduced by \citet{apruzzese_deep_2020}. We find that already after a single cycle of retraining on malicious perturbations, the classifier is hardened against future $5$-step exploits.} 
\label{fig:ithard}
\end{figure*}

Defense against NFC blackbox attacks is traditionally achieved by hardening, i.e. by supervised training on known (or artificially generated) malicious perturbations. We now demonstrate that, within the Apruzzese et al. (2020) framework, under favourable conditions, NFCs can be hardened sufficiently in just one retraining step. This means that subsequent attacks using the same method will largely fail to deceive the hardened classifier within over the same flow sample distribution. We thus present empirical evidence that, under certain conditions, blackbox MDEA dynamics, can indeed converge to a no-attack fixed point (see Figure \ref{fig:ithard}). However, within the MDEA framework, such fixed points are clearly unlikely to be stable as the attacker can choose to increase perturbation sizes at no cost.

\subsection{CyberMARL}

\begin{figure*}[h]
\begin{center}
\includegraphics[width=0.85\linewidth]{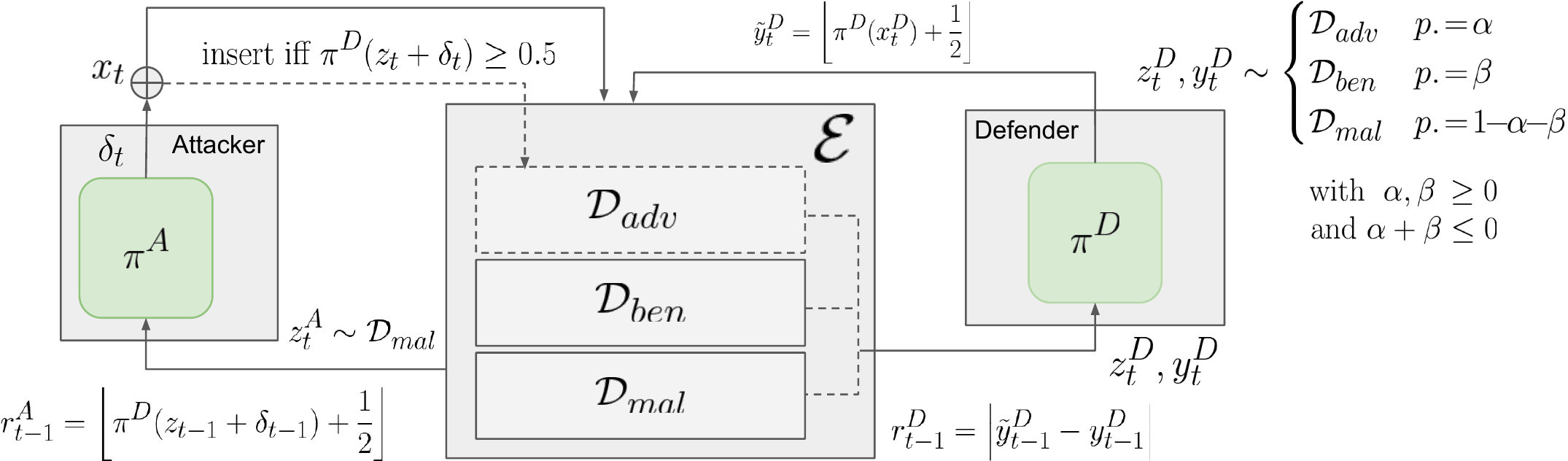}
\end{center}
\caption{A schematic overview of the \methodtwo~setting. Example network architectures for both attackers and defenders may be found in Appendix \ref{app:cyber}} 
\label{fig:cybermarl}
\end{figure*}

To investigate the dynamics of blackbox ADEA attacks, we now introduce a novel adversarial multi-agent game setting, \methodtwo, that allows to model the temporal co-evolutionary dynamics of both network attackers and defenders explicitly. In \methodtwo~ (see Figure \ref{fig:cybermarl}), attackers and defenders perform online training of their adversarial perturbation generators, and NFC networks, respectively. Attackers are rewarded if their generated perturbations evade the current defender policy, while defenders are rewarded if they classify adversarial perturbations correctly. In order to avoid runaway feedback effects, \methodtwo~slows the coupling of attacker and defender through the use of a shared adversarial sample buffer. Optionally, we additionally assume that all agents have access to a fixed NFC that has been trained on a ground truth distribution apriori. In each turn, both attacker and defender independently sample a statistical flow feature from the environment: The attacker receives a malicious sample for which she is to generate a suitable adversarial perturbation. In contrast, the defender randomly receives either a malicious, benign or adversarial sample from the environment and needs to correctly classify it as malicious or benign. 

In \methodtwo, the environment $\mathcal{E}$ stores all successful adversarial samples generated by the attacker in a buffer $\mathcal{D}_{adv}$. Otherwise, the attacker follows an identical process as in \methodone, with the subtle difference that not $C$, but dynamically updating defender policy $\pi^D$ has to be fooled.
\methodtwo~ also extends \methodone~ by adding a defender agent with policy $\pi^D$. At each episode start, the defender receives an i.i.d. sample $z^D$ from the environment which is sampled at probability $\alpha>0$ from $\mathcal{D}_{adv}$ \footnote{iff $\mathcal{D}_{adv}$ is not empty, else $\alpha$ is set to effective zero until it has been filled with at least one sample.}, instead from $\mathcal{D}_{ben}$ with probability $\beta>0,\ \alpha+\beta\leq 1$, and from $\mathcal{D}$ otherwise. The defender then receives a reward of $+1$ if it classifies $z^D$ correctly into either malicious or benign, i.e. iff $\lceil\pi^D(z^D)+\frac{1}{2}\rceil-y^D= 0$, and $0$ otherwise. 

Both attacker and defender policies are trained using the MADDPG framework (see section\ref{sec:maddpg}) under CTDE (see section \ref{sec:ctde}). Note that, instead of conditioning on the full state of the environment, we approximate $s_t$ (see section \ref{sec:maddpg}) by the union of agent observations $z_t^A$ and $z_t^D$. 

Overall, \methodtwo~can be summarised as a two-player adversarial game. Importantly, this game is not zero-sum in the conventional sense: rewards for attacker and defender are only weakly coupled through the the adversarial sample buffer $\mathcal{D}_{adv}$, and this coupling takes at least one timestep. This means that the defender policy may have changed in the meantime and an adversarial sample that worked at generation time may happen not to work anymore when sampled by the defender. In addition, \methodtwo~ can easily be extended with cooperative reward structures that incentivise both attackers and defenders to simulate the interest of both to keep network operations intact.

We let \methodtwo's policy network take the prediction score of the fixed classifier $C$ as additional input. To ensure that $\pi^D$ is initialised close to $C$, we employ a skip connection in $\pi^D$'s network architecture. Each fully connected layer in both actors and critics has $64$ activations and we employ the ReLU activation function. We choose $\gamma=0.95$ and a learning rate of $10^{-2}$.

\begin{figure}[h]
\centering
\includegraphics[width=0.5\linewidth]{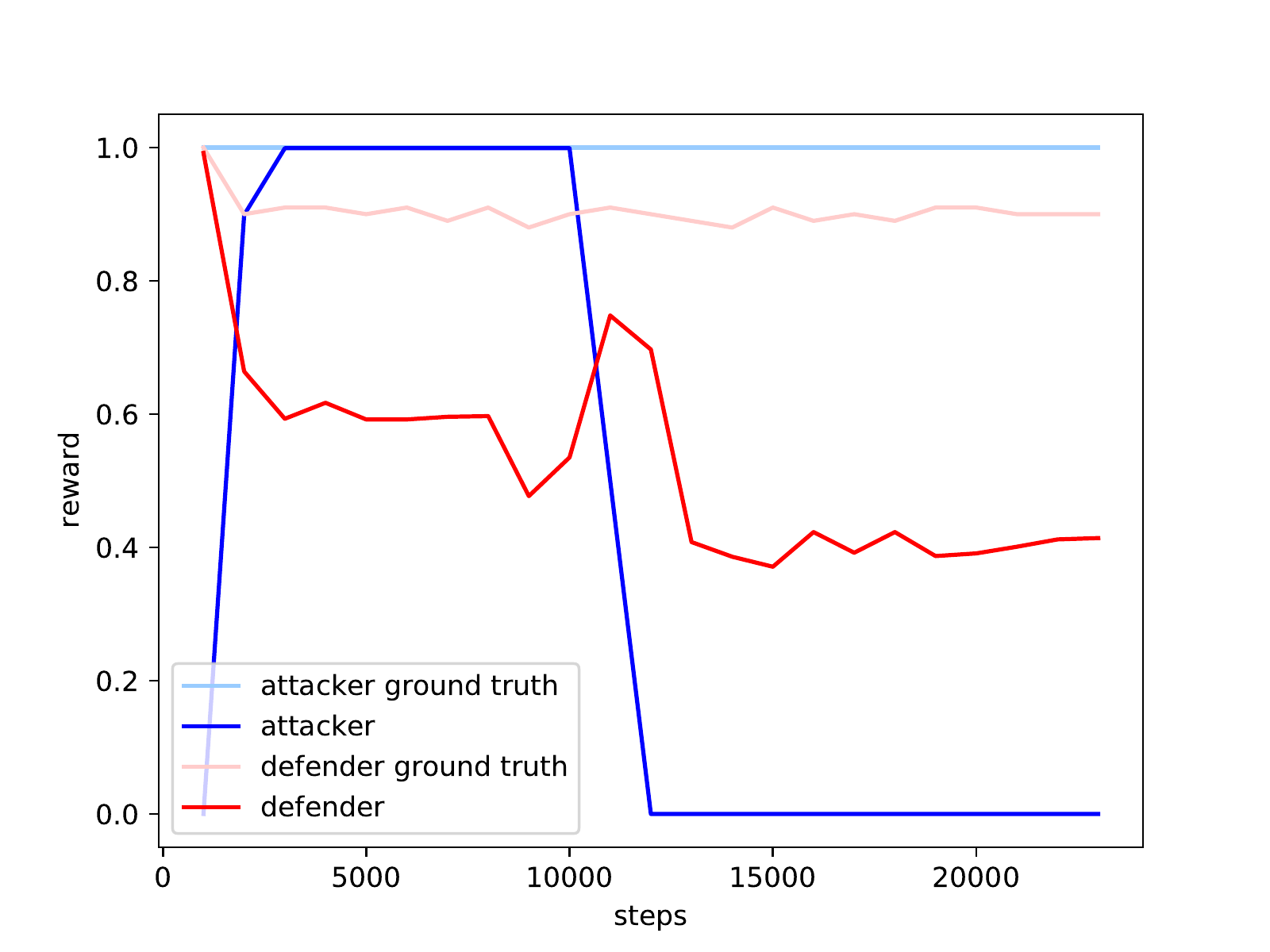}
\caption{Spontaneous reward transition in \methodtwo~framework for $\alpha=0.1,\ \beta=0.4$. The defender starts off with considerable skill. The attacker quickly learns how to reliably fool the defender, leading the defender's skill to drop slightly. A sudden transition renders the attacker entirely unskilled, with the defender temporarily regaining some skill, only to then drop to ca. $0.4$. The ground truth baselines indicates what rewards both agents would receive if the defender's policy was equal to the fixed classifier $C$. We see that the attacker always perfectly fools $C$. The evidence suggests that the continuous hardening of $\pi^D$ triggers a catastrophic transition point for the attacker.}
\label{fig:transition2}
\end{figure}

As a case study, we investigate a spontaneous reward transition in the \methodtwo~framework for $\alpha=0.1,\ \beta=0.4$ (see Figure \ref{fig:transition2}). The defender starts off with considerable skill. The attacker quickly learns how to reliably fool the defender, leading the defender's skill to drop slightly. A sudden transition renders the attacker entirely unskilled, with the defender temporarily regaining some skill, only to then drop to ca. $0.4$. The ground truth baselines indicates what rewards both agents would receive if the defender's policy was equal to the fixed classifier $C$. We see that the attacker always perfectly fools $C$. The evidence suggests that the continuous hardening of $\pi^D$ triggers a catastrophic transition point for the attacker.
However, it remains unclear why the defender stabilises at a reward of around $0.4$ specifically. The resulting collapse in defender performance may indicate that the policy network initialisation is unstable. This could be potentially alleviated by replacing the skip connection by a supervised pre-training step.

\section{Fixed Points of Attacker-Defender Co-Evolution}
\label{sec:white}

\subsection{Is there empirical evidence of attacker-defender co-evolution in Locked Shields?}

To gain some insight on possible real-world evidence of attacker-defender coevolution, we investigate flow classification in a real-world historic dataset from Locked Shields exercises (see Section \ref{sec:back}). While neither strategies of Team Red and Team Blue are publicly known, this paper uses a dataset of $>100M$ recorded package flow features from the years $2017$ to $2019$. Using a ground truth of knowingly infiltrated external nodes, this dataset is labeled for C\&C flows depending on whether a package flow involved a knowingly infected external node. This labeling process is of course not exact and omits potential package flows that are relayed through infected internal nodes.

To assess the difficulty of separating malicious from benign flows, we train a gradient-boosting random forest classifier (\textit{LightGBM}) \citep{ke_lightgbm_2017} on a class-balanced dataset of $1$m samples, using all $80$ features provided. We find that our classifier achieves an accuracy of more than $98\%$ when trained and evaluated on samples from both $2017$ and $2018$. However, accuracy decreases, and, in particular, false positive rates significantly increase if classifiers trained on a particular year are validated on another year (see Table \ref{tab:forecasts}). This implies that not only malicious flows changed slightly between consecutive Locked Shields exercises, but also the characteristics of the benign background flows. As we do not have sufficient information about changes in attacker-defender methodologies over time, however, we cannot establish any causal relationship for these changes, hence cannot definitely ascribe them to co-evolutionary dynamics.

\begin{table}[h]
  \centering
  \small
\begin{minipage}[c]{0.48\textwidth}
\centering
    \begin{tabular}{@{\extracolsep{5pt}} lccc}
      \toprule
    Train \textbackslash  Val  & 2017 & 2018 & all \\ 
      \midrule
    2017 & 0.9851 & 0.9680 & 0.9785 \\
    2018 & 0.9708 & 0.9956 & 0.9799\\
    all & 0.9799 & 0.9945 & 0.9847 \\
      \bottomrule
    \end{tabular}
\captionof{table}{Predicting malicious network traffic}
\end{minipage}
\begin{minipage}[c]{0.48\textwidth}
\centering
    \begin{tabular}{@{\extracolsep{5pt}} lccc}
      \toprule
    Train \textbackslash  Val  & 2017 & 2018 \\ 
      \midrule
    2017 & $0.9(3)$ & $0.5(6)$\\
    2018 & $0.9(5)$ & $0.1(0.8)$ \\
      \bottomrule
    \end{tabular}
\captionof{table}{False negatives (positives), in \%}
\end{minipage}
\caption{ROC under AUC for binary LightGBM classifiers trained and evaluated on pairs of $10^6$ class-balanced Locked Shield labels (top 20). Elevated false positive (negative) rates for cross-annual evaluation indicate that distributions of both malicious and benign network traffic changed significantly between exercises.
  }
  \label{tab:forecasts}
\end{table}

\subsection{Whitelisting: A Possible Fixed Point of Attacker-Defender Co-evolution?}

\begin{figure}[h]
\centering
        \centering
        \includegraphics[width=0.6\linewidth]{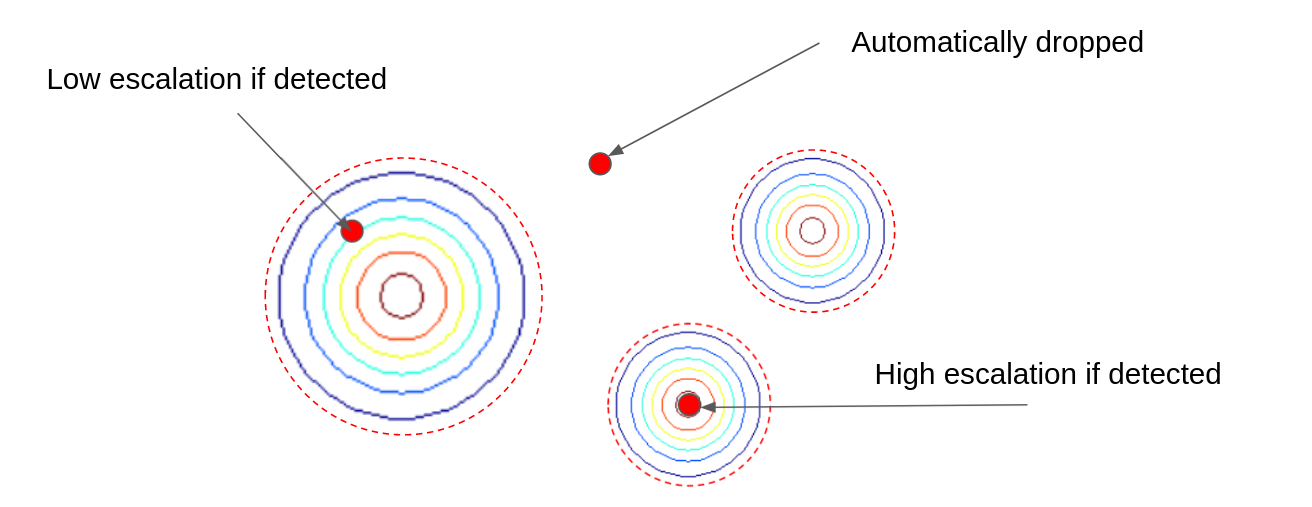}
        \caption{A schematic depiction of \textit{Whitelisting}. The concentric circles denote different whitelisting 'corridors', i.e. statistical flow feature subspaces outside of which traffic is automatically blocked. The red dots denote different types of adversarial perturbations. We assume that adversarial perturbations, if falling closer to the centre of each whitelisted corridor, are more likely to cause major system disruption if detected due to the increased amount of background traffic in these areas. }
        \label{fig:whitelisting}
\end{figure}

We argue that the adoption of automated defense systems (AI-NIDS) will stimulate the development of automated attack systems, and in turn dramatically shorten attacker-defender co-evolutionary cycles. This poses the question of whether attacker-defender co-evolution has any natural fixed points, and whether the system will eventually converge to these. We hypothesise that one possible fixed point could be a setting we refer to as \textit{Whitelisting Hell}, in which defenders restrict flow feature space to a dynamic distribution of narrow subspaces, outside of which traffic is dropped automatically. This would counteract the attacker's capabilities of traversing freely in flow feature space, but limit the networks ad-hoc operability. \textit{Whitelisting Hell} (see Figure \ref{fig:whitelisting}) therefore poses a fundamental tradeoff between security, and operational flexibility that is already encountered in IoT security research. 

Even in \textit{Whitelisting Hell}, attackers may learn successful ADEA schemes. We suggest that in such cases, one may consider the factorised problem of the attacker first needing to predict which SFF subspaces are whitelisted ahead of time, and then learn to generate adequate adversarial perturbations separately for each subspace. As in \methodtwo, attackers need to ensure that the perturbations generated are unlikely to trigger major security escalations if detected by NIDS infrastructure (particularly behavioural anomaly detection not based on NFC approaches), while keeping botnet operability intact.

\subsection{Attacker-Defender Co-evolution: A Continual Learning Perspective}

\begin{figure}[h]
\centering
\begin{minipage}{1.0\textwidth}
    \begin{minipage}{0.24\textwidth}
            \includegraphics[width=1.0\linewidth]{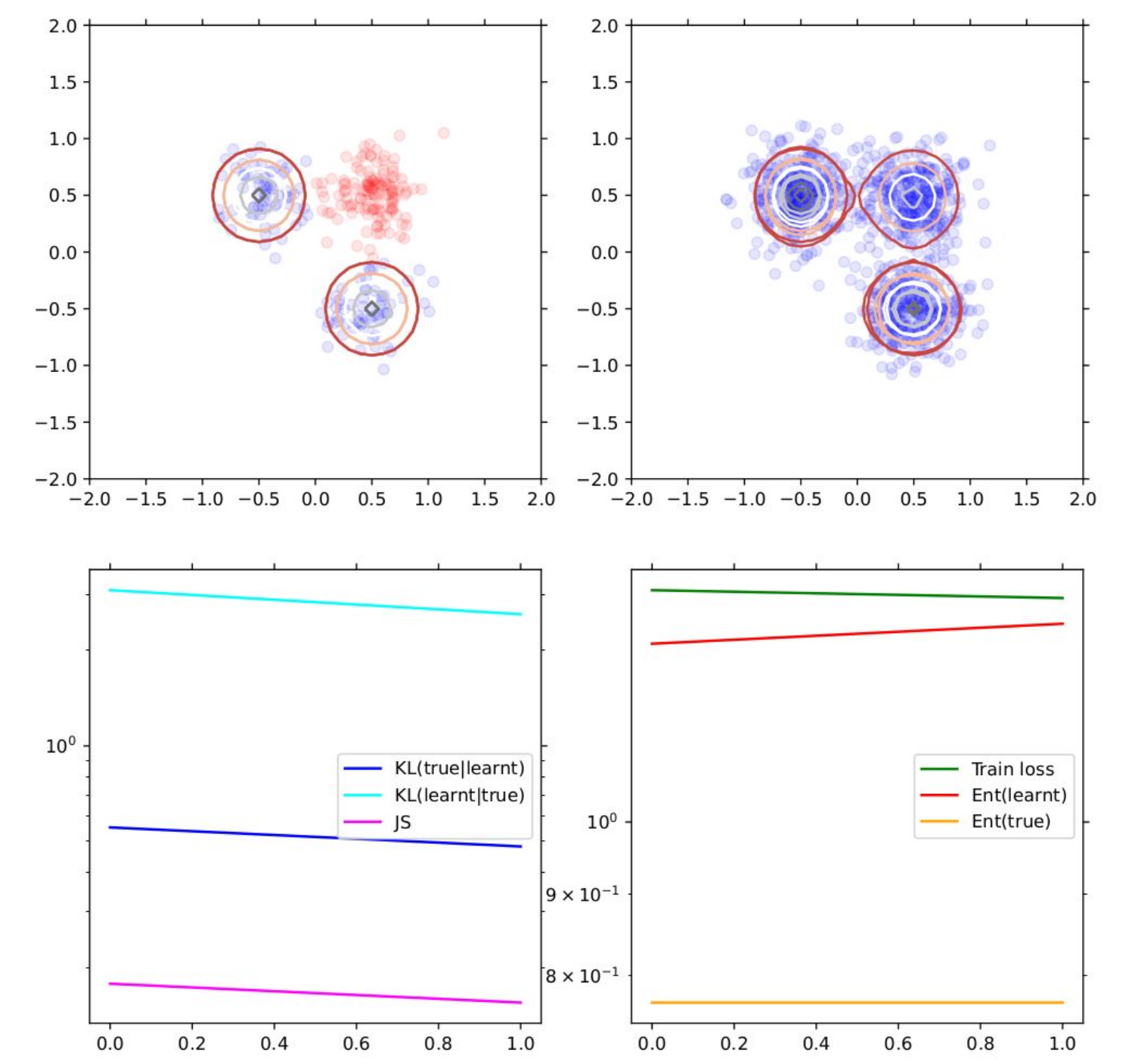}
    \end{minipage}
    \begin{minipage}{0.24\textwidth}
            \includegraphics[width=1.0\linewidth]{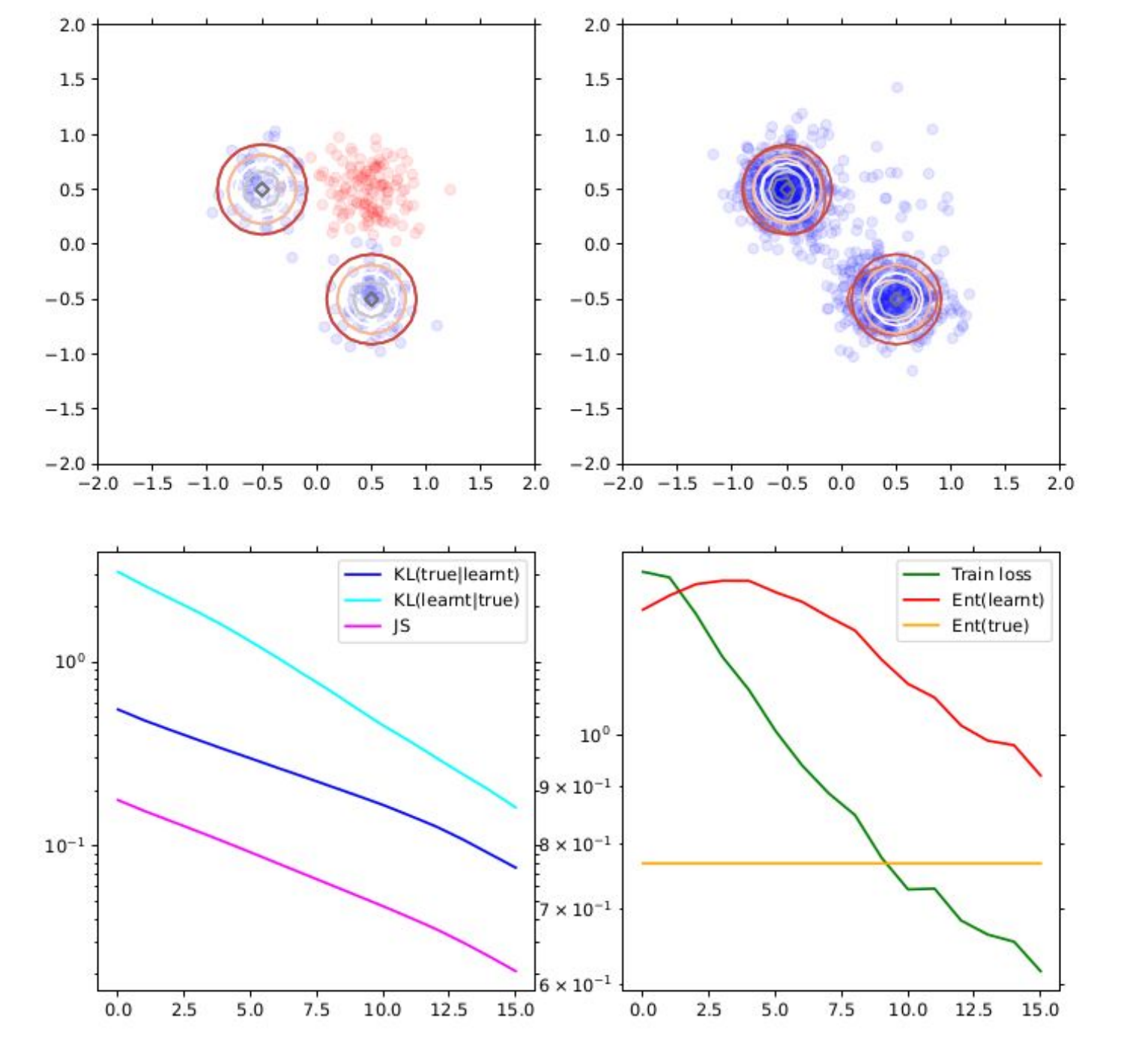}
    \end{minipage}
    \begin{minipage}{0.24\textwidth}
            \includegraphics[width=1.0\linewidth]{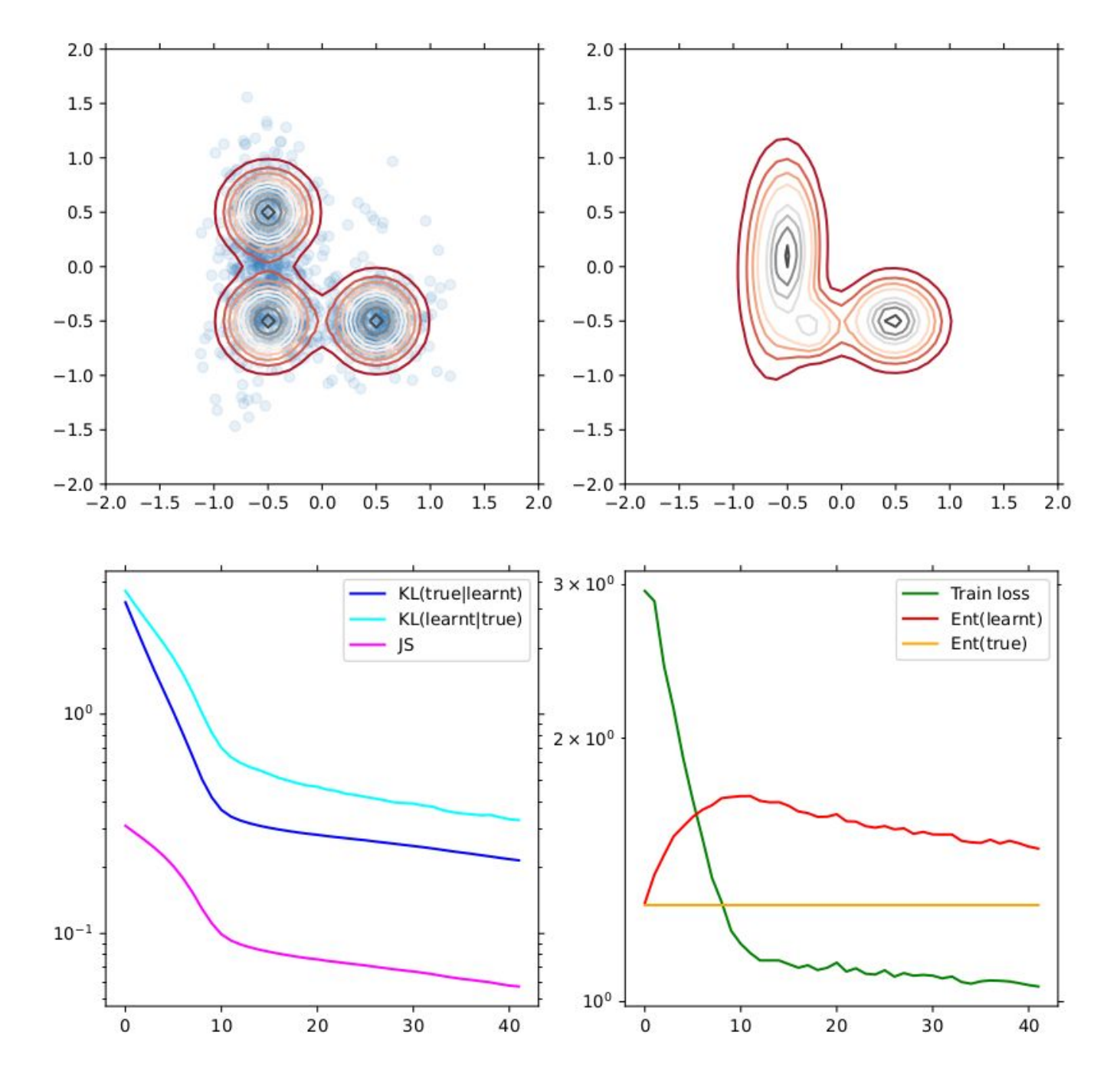}
    \end{minipage}
    \begin{minipage}{0.24\textwidth}
            \includegraphics[width=1.0\linewidth]{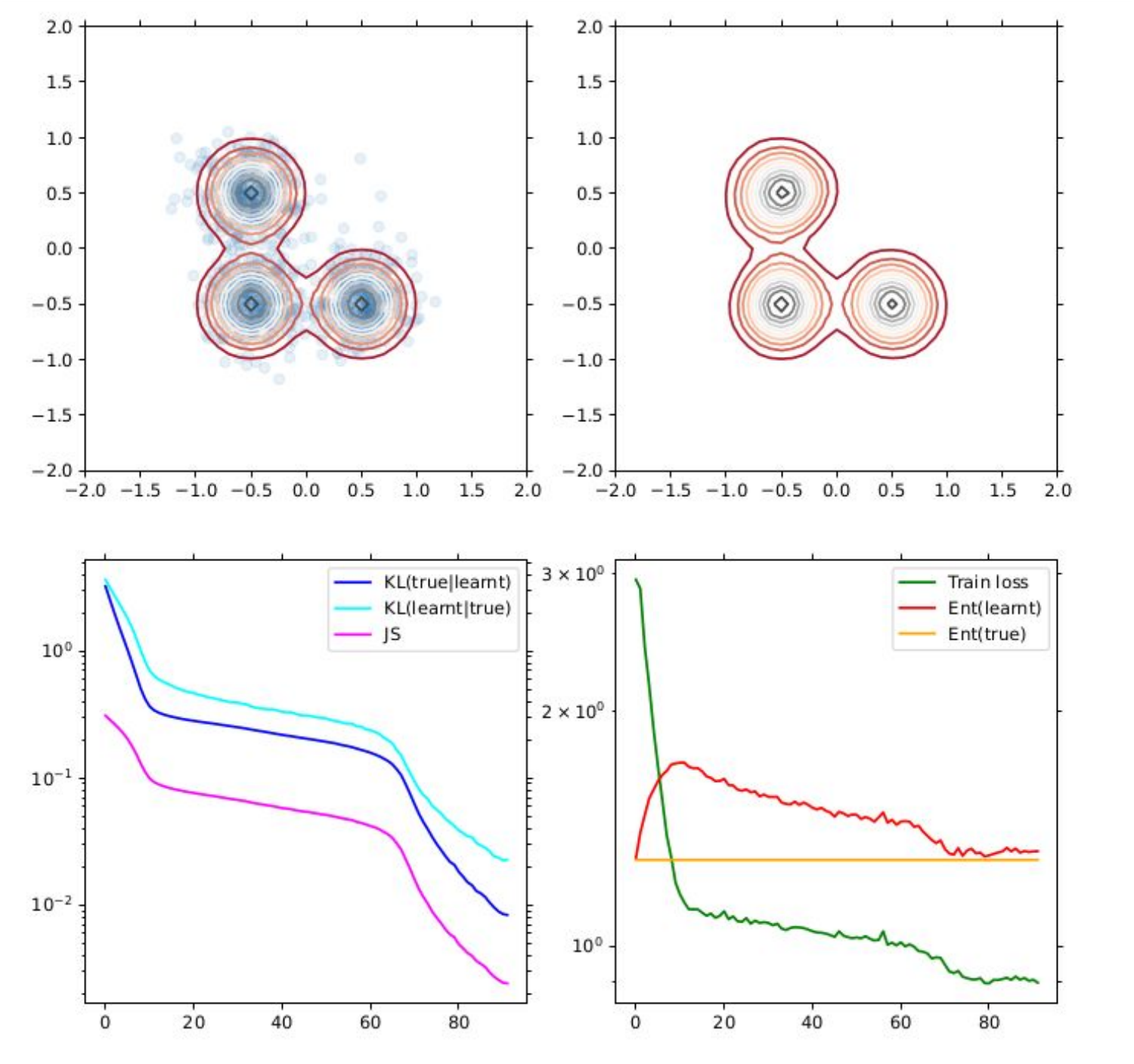}
    \end{minipage}
\end{minipage}\\
\begin{minipage}{1.0\textwidth}
    \begin{minipage}{0.24\textwidth}
            \includegraphics[width=1.0\linewidth]{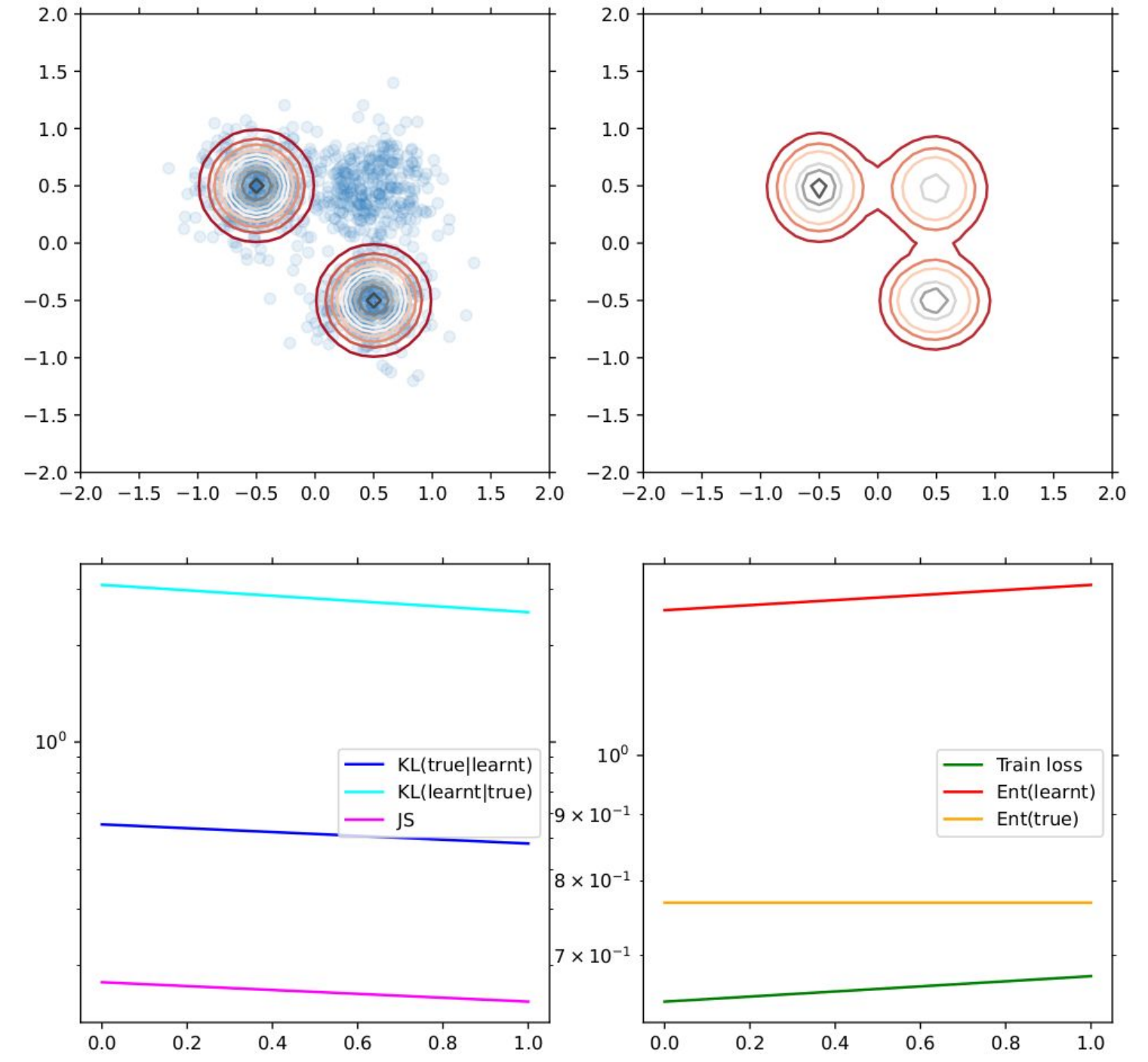}
    \end{minipage}
    \begin{minipage}{0.24\textwidth}
            \includegraphics[width=1.0\linewidth]{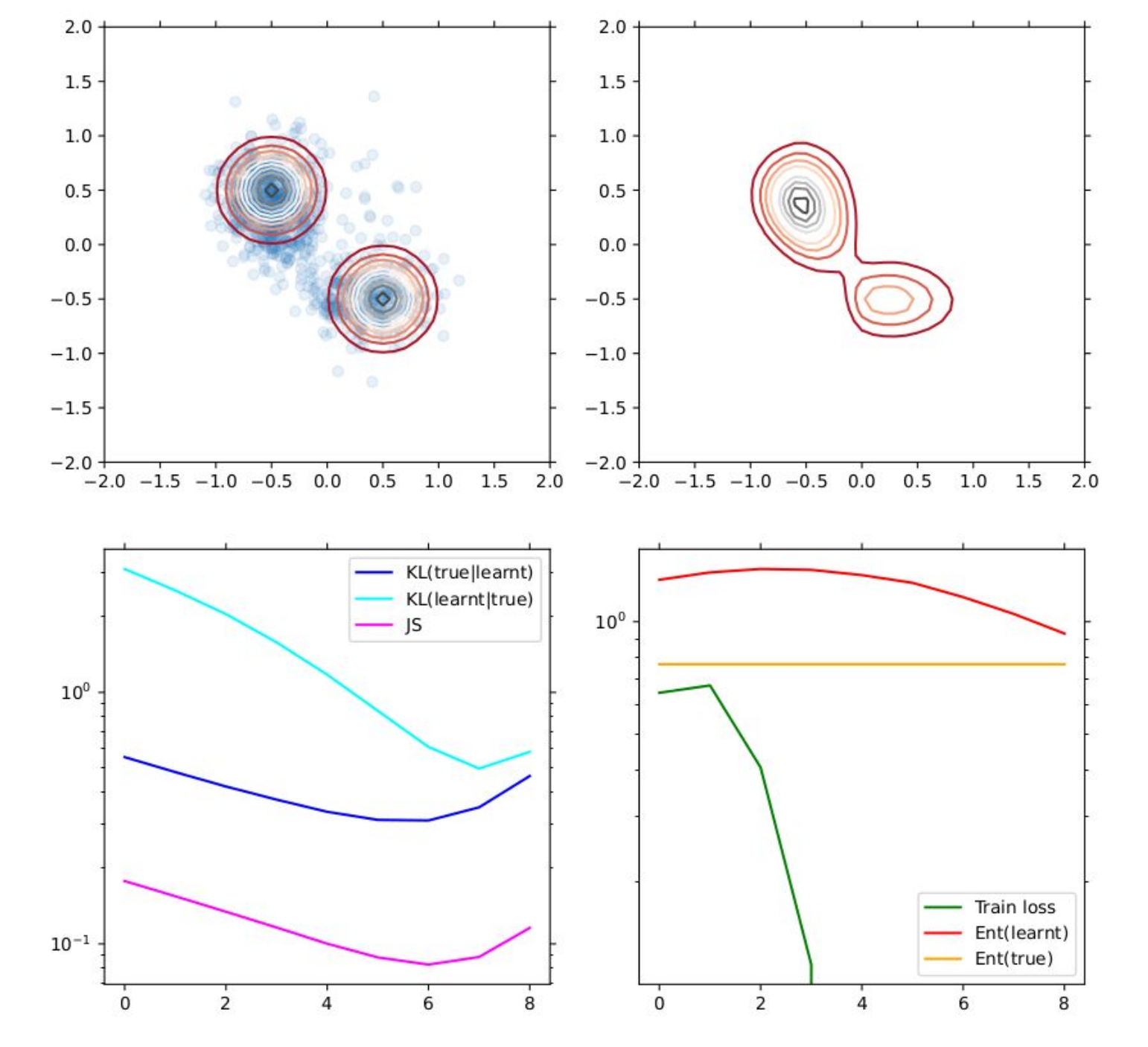}
    \end{minipage}
    \begin{minipage}{0.24\textwidth}
            \includegraphics[width=1.0\linewidth]{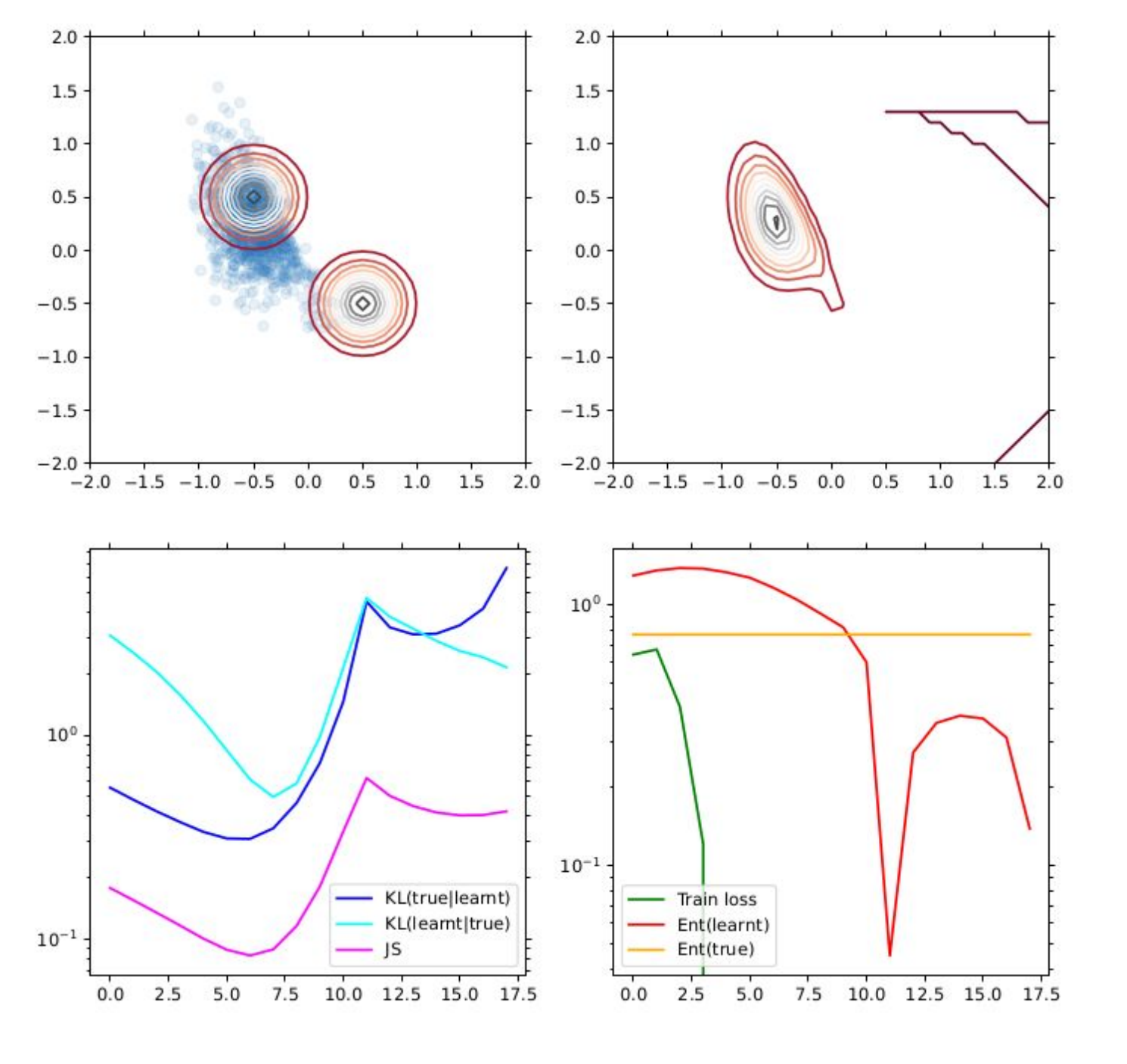}
    \end{minipage}
    \begin{minipage}{0.24\textwidth}
            \includegraphics[width=1.0\linewidth]{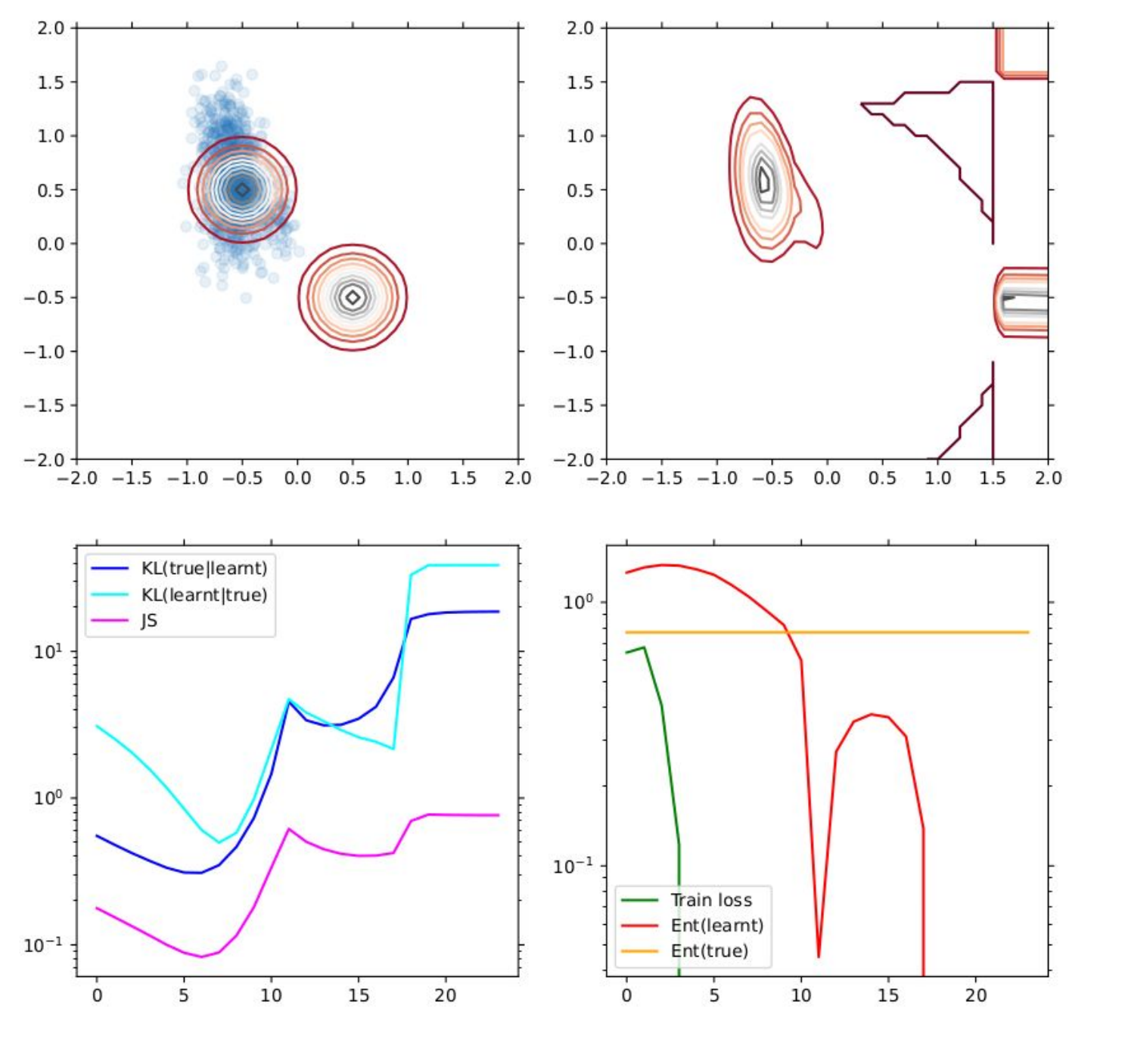}
    \end{minipage}
\end{minipage}\\
\caption{Illustration of the learning process of an autoregressive generative model when unlearning a mode of the distribution (top left and center left), and subsequently adding a distributional mode (top center right and right). Bottom left to right: Unlike at the top, this random initialisation results in an unstable adaption process when sampling distribution has top right mode removed, highlighting the need for a continual learning approach to adversarial perturbation generation in temporally-extended general sum games.}
\label{fig:forget}
\vspace{-0.5cm}
\end{figure}

Whether in \textit{Whitelisting Hell}, or under the more general ADEA conditions explored by \methodtwo, possible attack strategies may seek to exploit a weakness of supervised classifiers: Neural networks, as well as other classifiers, suffer from catastrophic forgetting. Catastrophic forgetting occurs when a network's weights do no longer reflect the gradient updates associated with samples that it has not seen in training over an extended period of time. In \methodtwo~ settings, attackers might exploit defender NFCs by confining itself to perturbation subspaces for a prolonged period of time, before suddenly switching to a different perturbation subspace which the NFC has not been trained on for an extended period of time. In Figure \ref{fig:forget}, we illustrate various learning instabilities when generative networks are trained on multi-modal data whose modes suddenly change during the learning process.

As an antidote to catastrophic forgetting, training samples could be exhaustively stored and the classifier be retrained continuously, however, this may not be possible given traffic volumes, as well as real-time constraints and computation budgets even of large-scale AI-NIDS systems. We suggest and empirically illustrate that NFC defenses in future AI-NIDS may benefit from \textit{continual learning} \citep{parisi_continual_2019} techniques. In particular, we suggest that defender networks should employ a reinforcement learning variant of \cite[A-GEM]{chaudhry_efficient_2019}, which is based on retaining a small set of characteristics samples in a replay buffer that can be used for retraining.

\begin{figure}[h]
\centering
        \centering
        \includegraphics[width=0.7\linewidth]{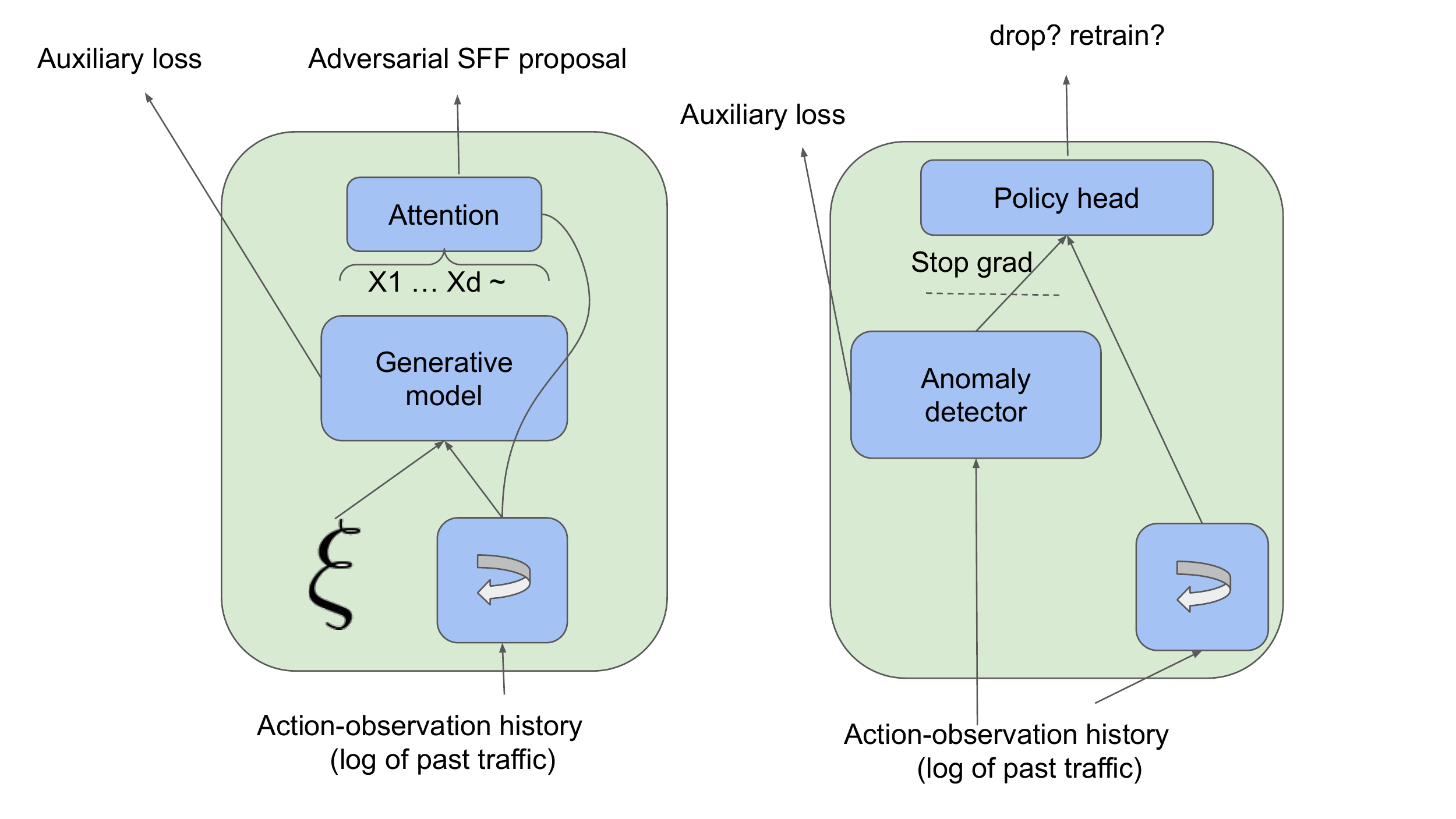}
        \caption{Suggested network architectures for \methodtwo~in the continual learning setting. Left: attacker policy network, Right: defender policy network.}
\label{fig:cont}
\end{figure}

\pagebreak

\section{Related Work}

There exists a large body of literature on \textit{adversarial attacks} against neural network classifiers and we do not guarantee completeness of our literature review \footnote{If you believe that a particular work needs to be included here, please contact the authors.}. Early methods include the fast gradient sign method (FSGM) \citep{goodfellow_explaining_2015}, as well as several optimization-based methods \citep{carlini_towards_2017,liu_delving_2017,xiao_spatially_2018,eykholt_robust_2018}. Traditional work on optimal evasion attacks, on and hardening of, non-neural network network flow classifiers focuses on the generation of adversarial samples that minimise the $L_p$-norm to real samples \citep{merkli_evaluating_2020}. 

Reinforcement learning is a learning setting in which an agent interacts with an environment trying to optimise a scalar reward Deep reinforcement learning (DRL), which employs DNNs as function approximators, dates back to a seminal paper by \citep{mnih_human-level_2015}. Since then, single-agent DRL has found a number of applications in cybersecurity \citep{nguyen_deep_2020}. A number of deep multi-agent reinforcement learning for partially-observable stochastic games have been proposed, ranging from model-free actor-critic variants \citep{openai_dota_2019, lowe_multi-agent_2020} to Monte-Carlo tree search with self-play \citep{silver_mastering_2017, schrittwieser_mastering_2020}. 

A number of recent papers study certain cyber security scenarios as temporally-extended adversarial games: \citet{eghtesad_adversarial_2020} seek to thwart attacks on cyber networks by continuously changing their attack surface, ie. the configuration of hosts and network topology, using deep reinforcement learning. \citet{ye_differentially_2021} propose an approach in which a defender adopts differential privacy mechanisms to strategically change the number of systems and obfuscate the configurations of systems, while an attacker adopts a Bayesian inference approach to infer the real configurations of network systems. None of these works explicitly treats botnet infiltration and detection scenarios.

Work concurrent to this paper uses off-policy double Q-learning in order to harden a network flow classifier using adversarial sample generation guided by a sparse reward feedback signal \citep{apruzzese_deep_2020}. Contrary to our framework, they do not consider simultaneous evolution of both attacker and defender. In addition, \citet{apruzzese_deep_2020}'s approach only allows for sparse and \textit{discrete} perturbations to be generated, while our approach based on DDPG allows for arbitrary perturbations to be generated. In particular, the latter crucially allows us to learn perturbations that involve changes to multiple features simultaneously.


\section{Conclusions and Future Work}

In this whitepaper, we are charting a possible future in which the automation of cyber defense systems (such as AI-NIDS) has accelerated the traditional co-evolutionary cycle between attackers and defenders by orders of magnitude. Consequently, we propose that in such a world, cyber defense research might itself need to depart from finding responses to the latest threats in a reactionary fashion, and instead focus on rigorously studying the co-evolutionary system properties themselves, including their fixed points. At the same time, we have challenged a central tenet in the study optimal evasion attacks: namely that attackers do not necessarily incur a cost proportional to the size of the adversarial perturbation, but rather can, even today, generate arbitrary ones at negligible cost. 

Translating all of these insights to the setting of botnet detection and mitigation, we first introduce \methodone, an efficient reinforcement-learning based blackbox adversarial attack generator that can generate both discrete and continuous perturbations of arbitrary size within a single model episode step. We empirically show that \methodone~ can also be efficiently used in conventional minimal evasion distance settings, although a detailed benchmarking against \cite{apruzzese_deep_2020}'s approach is left for future work, as are extensions with probabilistic policies. 

With \methodtwo, we introduce a novel setting that allows attacker-defender dynamics to be studied as a temporally-extended general-sum game. While our empirical investigation of this setting are preliminary, \methodtwo~ opens up a whole new research program for the study of fixed point dynamics of automated attacker-defender systems. 

We also hypothesise that a fixed point of attacker-defender co-evolution could be a setting with extensive use of network traffic \textit{whitelisting} and argue that classifiers of future AI-NIDS will require techniques from the continual learning community in order to avoid forced catastrophic forgeting.





\section*{Acknowledgements}
We thank Vincent Lenders, Giorgio Tresoldi, Luca Gambazzi, and Klaudia Krawiecka for helpful discussions. Christian Schroeder de Witt is generously funded by Cyber Defence Campus, armasuisse Science and Technology, Switzerland. This project has received funding from the European Research Council under the European Union’s Horizon 2020 research and innovation programme (grant agreement number 637713). The experiments were made possible by a generous equipment grant from NVIDIA.


\bibliography{iclr2022_conference}
\bibliographystyle{iclr2022_conference}

\pagebreak
\appendix
\pagebreak
\section{Appendix}
\raggedbottom
\subsection{Data rewriting}
\label{app:data}

The large number of samples contained in the Locked Shields dataset and resulting CSV-file sizes (>100GB) are prohibitively slow to access, even using distributed frameworks such as Dask \citep{dask}, for both exploratory analysis, as well as for model training. In contrast to storage formats requiring \textit{read} system calls, including HDF5\footnote{\url{https://portal.hdfgroup.org/display/HDF5/HDF5}(2021)}, Zarr\footnote{\url{https://zarr.readthedocs.io/en/stable/} (2021)} or xarray\footnote{\url{http://xarray.pydata.org/en/stable/} (2021)}, memory-mapped files use the \textit{mmap} system call to map physical disk space directly to virtual process memory, enabling the use of \textit{lazy} OS demand paging and circumventing the kernel buffer. This makes memmaps particularly efficient for random access patterns, such as commonly found during model training.


We therefore rewrite each data feature into an individual \textit{numpy memmap} file. As we choose the primary axis to be ordered by time, access to specific time indices can be efficiently handled using numpy's \textit{searchsorted}.

\begin{table}[h]
\centering\small
\begin{tabular}{@{\extracolsep{5pt}}| l|c|c|c|c|}
\bottomrule
  Feature ID & Mean0 & Mean1 & Std0 & Std1\\\toprule\bottomrule
    Protocol & 12.4 & 6.0 & 5.51 & 9.6E-2 \\\hline
    dstIntExt & 0.13 & 0.10 & 0.34  & 0.05  \\\hline
    Active Mean & 2E6 & 1E6 & 3E4 & 3E5\\\hline
    Init Fwd Win Byts & 3E9 & 8E5 & 2E9 & 6E7 \\\hline
    FIN Flag Cnt & 0.18 & 0.97 & 0.38 & 0.17 \\\hline
    Bwd Pkt Len Min & 34.7 & 0.0 & 101.2 & 0.2  \\\hline
    Flow Pkts/s & 2E4 & 2E3 & 1E5 & 4E4 \\\hline
    Fwd IAT Max & 1E6 & 8E4 & 2E6 & 6E5 \\\hline
    Fwd IAT Min & 2E5 & 2E4 & 1E6 & 2E5 \\\hline 
    Subflow Fwd Pkts & 8.6 & 11.5 & 141.2 & 7.4 \\\hline
    Flow IAT Max & 1.5E6 & 8E5 & 2.7E6 & 6E5 \\\hline
    Fwd IAT Tot & 2E6 & 1E5 & 3E6 & 9E5  \\\hline
    Subflow Bwd Pkts & 3.4 & 6.0 & 196.0 & 7.0 \\\hline
    Subflow Fwd Byts & 3.83E & 1.5E3 & 1E5 & 7.6E3 \\\hline
    Bwd Header Len & 63 & 131 & 4.4E3 & 141 \\\hline
    Tot Bwd Pkts & 3.4 & 6.0 & 196 & 7 \\\hline
    Fwd Pkt Len Std & 26 & 197 & 81 & 44 \\\hline 
    Fwd Seg Size Min & 15.6 & 20.3 & 10.1 & 1.6 \\\hline
    Bwd Pkt Len Std & 20 & 86 & 74 & 36 \\\hline 
    Bwd IAT Mean & 1E5 & 1E4 & 5E5 & 1E5 \\
  \toprule
\end{tabular}
\caption{Two-column list of the $20$ statistical flow features (CICFlowMeter) used across all empirical evaluations in this paper. Means and standard deviations are listed for both benign ($0$) and malicious ($0$) features. Note that even large differences in dataset statistics do not necessarily imply that individual samples are easily distinguished. Features were derived using feature importance scores after training on all $80$ features available using a LightGBM model.}
\label{table:feat}
\end{table}

\pagebreak
\subsection{RL training pathologies}
\label{app:path}

\begin{figure}[h]
\centering
        \centering
        \includegraphics[width=1.0\linewidth]{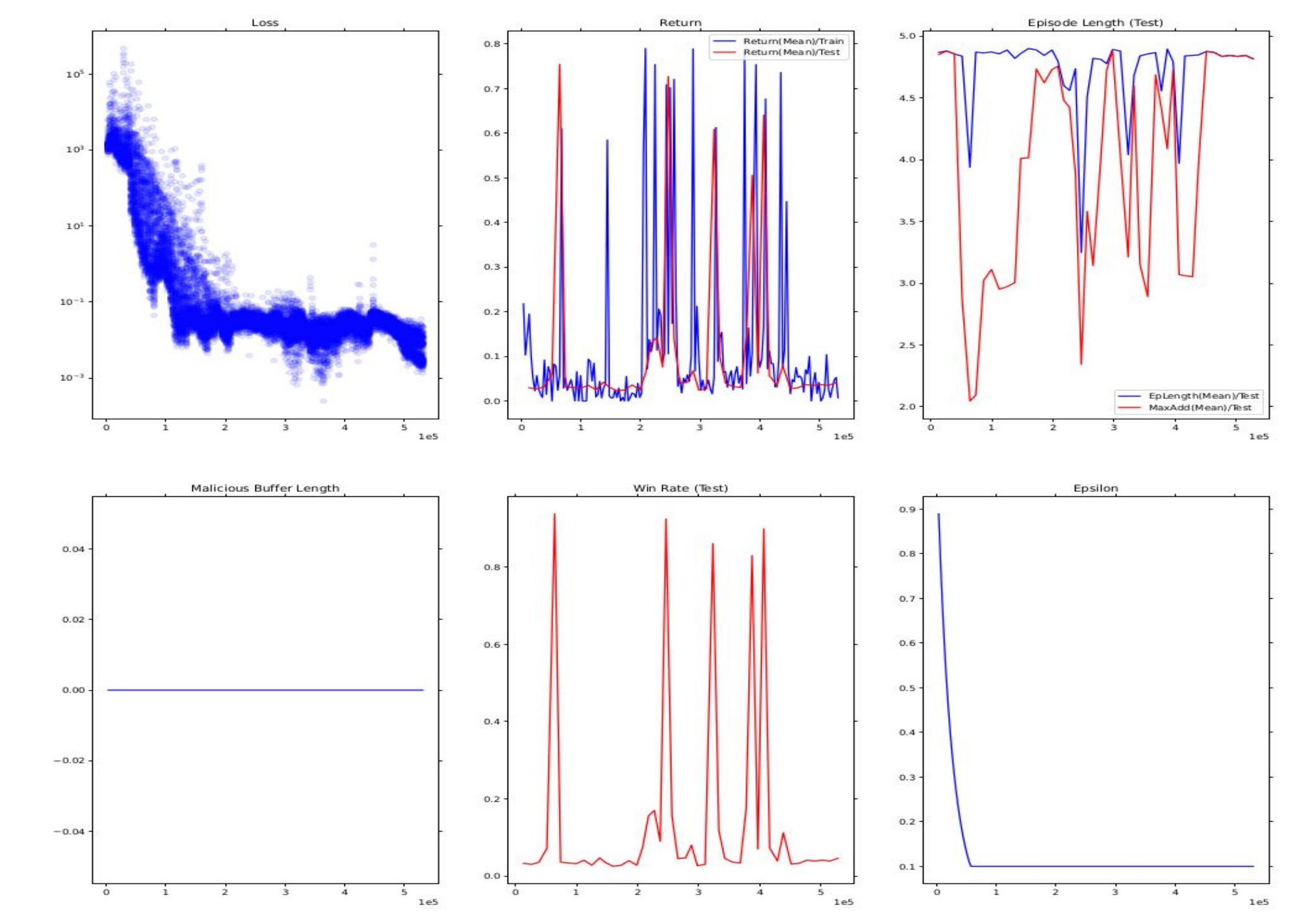}
        \caption{An illustration of common learning pathologies in RL-based blackbox attack training (here within the framework of \citet{apruzzese_deep_2020}): We erroneously chose not to condition our agents' observations on either timesteps (in cases with maximum timestep cutoffs) or the current perturbation delta (for cases in which a episodes are terminated once a maximum feature space distance to the original has been reached).}
\label{fig:rl_training_pathologies}
\end{figure}

\pagebreak
\subsection{Example policy network architectures}
\label{app:cyber}

\begin{figure*}[h]
\label{fig:networks}
    \centering
    \begin{minipage}{0.3\textwidth}
        \centering
        \includegraphics[ height=0.3\textheight]{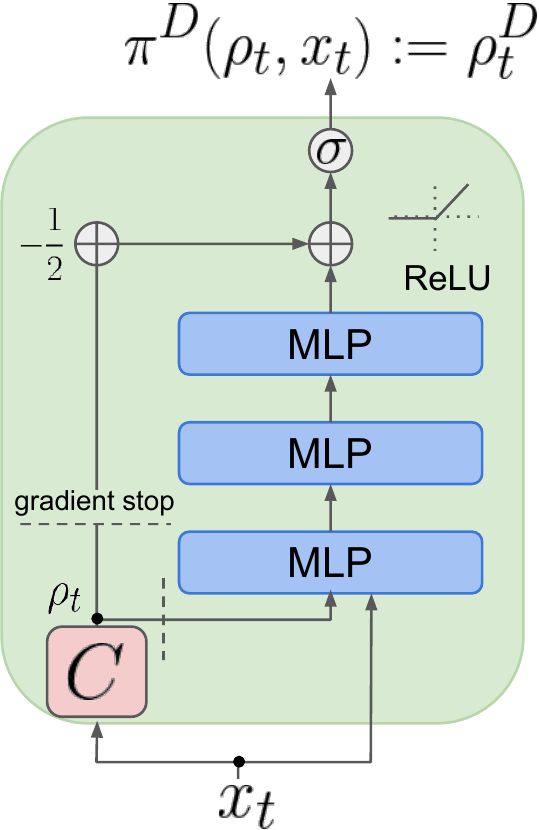}
        \caption{Defender policy\\ network architecture }
        \label{fig:marl_defender}
    \end{minipage}
    \begin{minipage}{0.57\textwidth}
        \centering
        \includegraphics[ height=0.3\textheight]{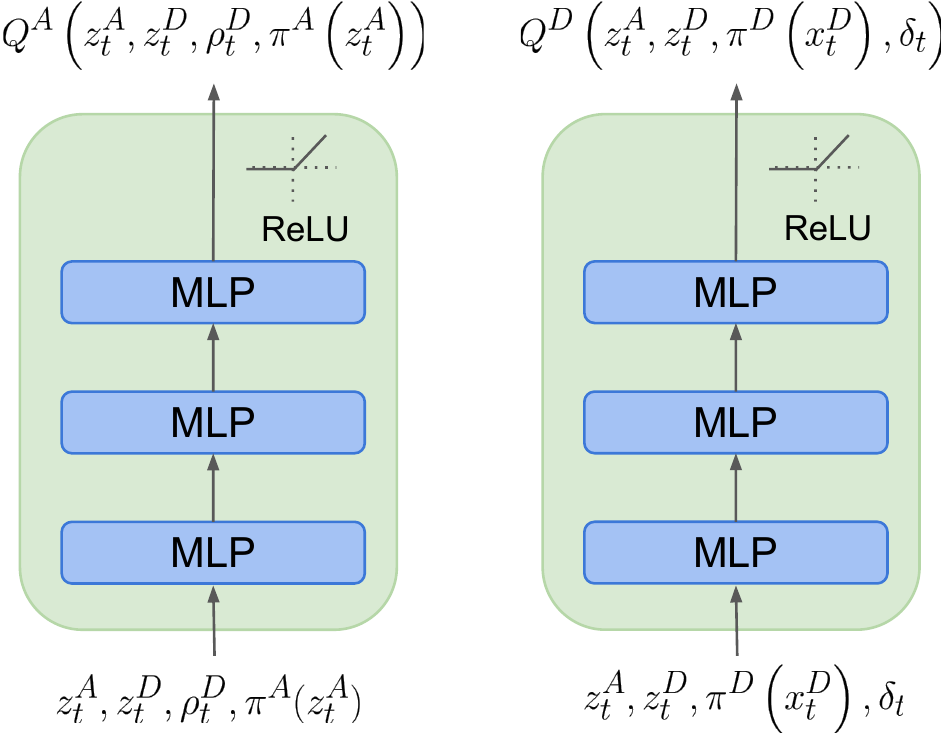}
        \caption{Critic network architectures for both attacker and defender (\methodtwo)}
        \label{fig:marl_critic}
    \end{minipage}    
\end{figure*}

\begin{figure}[h]
\label{fig:policy_networks}
\centering
    \begin{minipage}{0.18\textwidth}
        \centering
        \includegraphics[ height=0.2\textheight]{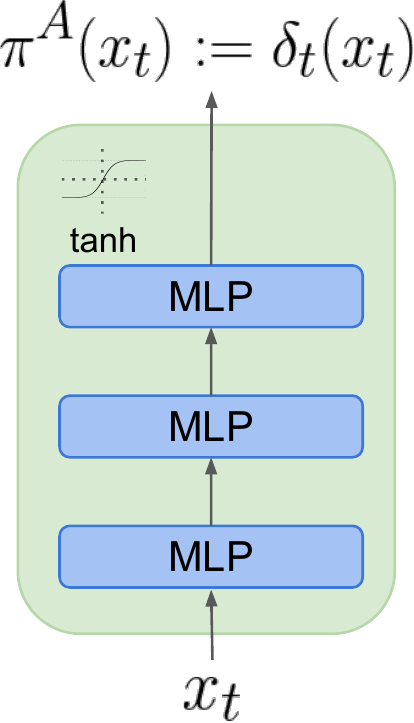}
        \caption{\quad\\(\methodone)\\ Attacker policy\\ network architecture}
        \label{fig:drl_policy}
    \end{minipage}
    \begin{minipage}{0.18\textwidth}
        \centering
        \includegraphics[ height=0.2\textheight]{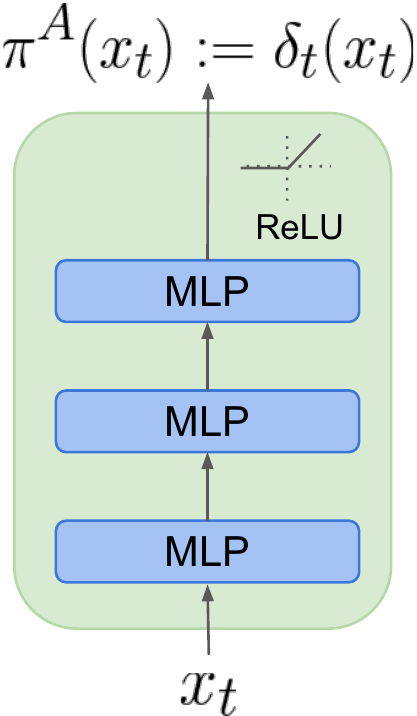}
        \caption{\quad\\(\methodtwo)\\ Attacker policy\\ network architecture}
        \label{fig:marl_attacker}
    \end{minipage}
\end{figure}

\end{document}